\documentclass[12pt,preprint]{aastex}
\usepackage{natbib}
\usepackage{textcomp}
\usepackage{amsmath}
\usepackage{subfigure}
\usepackage{gensymb}
\usepackage{url}
\usepackage{graphicx}
\shortauthors{Brantseg et al.}

\begin{document}
 \title{A Multi-Wavelength Look at the Young Plerionic Supernova Remnant 0540-69.3}
 \author{T. Brantseg$^1$, R. L. McEntaffer$^1$, L. M. Bozzetto$^2$, M. Filipovic$^2$, N. Grieves$^1$}
 \affil{$^1$Department of Physics and Astronomy, University of Iowa, Iowa City, IA 52242}
 \affil{$^2$School of Computing, Engineering, and Mathematics, University of Western Sydney, Penrith, New South Wales 2751}
 \email{thomas-brantseg@uiowa.edu}
 
\begin{abstract}
We present a study of the plerionic supernova remnant $0540-69.3$ in the LMC in X-ray, radio, optical, and infrared. We find that the shell of $0540-69.3$ is characterized in the X-ray by thermal nonequilibrium plasma with depleted Mg and Si abundances and a temperature of $kT \sim 0.7$ keV. This thermal emission is superimposed with synchrotron emission in several regions. Based on X-ray spectra and on morphological considerations in all surveyed wavebands, we conclude that the shell is expanding into a clumpy and highly inhomogeneous medium. In one region of the shell we find an overabundance of Ne, suggesting the presence of ejecta near the edge of the remnant. We also see evidence for reheating of material via a reverse shock originating from the interaction of the supernova blast wave with a particularly dense cloud in the surrounding medium. Finally, we perform the first detailed study of the "halo" region extending $1.2-2.2$ pc from the central pulsar.  We detect the presence of thermal and nonthermal spectral components but do not find evidence for mixing or ejecta.  We conclude that the thermal component is not a counterpart to similar optical and infrared halos and that it is most likely due to the projection of shell material along the line of sight.
\end{abstract}
 
\section{Introduction}
Supernova remnants (SNRs) are key to many important physical processes within galaxies, including creation and dispersal of heavy elements, creation of compact objects (neutron stars and black holes) and cosmic ray generation. SNRs can also serve as probes of the density and composition of the interstellar medium (ISM) surrounding the progenitor star, which is heated to X-ray emitting temperatures by the strong expanding shocks produced in the supernova explosion. Plerionic composite SNRs are a subclass of core-collapse SNRs that feature not only a normal SNR shell but also a synchrotron-emitting pulsar wind nebula (PWN), also called a plerion, surrounding the pulsar created in the supernova explosion. The PWN is powered by the conversion of the pulsar's spin-down luminosity into a shocked relativistic charged-particle wind, which interacts with the pulsar's strong magnetic field to produce synchrotron radiation. As the bubble of synchrotron-emitting material expands inside the SNR, it drives a shock outward into the surrounding medium, which is usually dominated by ejecta in young SNRs.

SNR $0540-69.3$ (hereafter 0540) is one of only five known plerionic composite SNRs in the Large Magellanic Cloud, alongside N206 \citep{WilliamsChu2005}, DEM L241 \citep{BambaUeno2006}, N157B \citep{WangGotthelf1998,DennerlHaberl2001,ChenWang2006}, and $0453-68.5$ \citep{Gaensler2003,McEntafferBrantseg2012}. Of these, 0540 is the youngest, although age estimates vary. \citet{KirshnerMorse1989} calculated an age of 760 years based on the velocity broadening of optical lines around the PWN, while \citet{SewardHarnden1984} calculated an characteristic age of the pulsar of 1600 years. The most commonly used age estimate is 1100 years \citep{Reynolds1985}, based on a calculation of the initial period from the break in the pulsar's broadband spectrum together with an estimate of the pulsar's braking index. 

The pulsars and PWNe in 0540 and the Crab Nebula share a long list of similarities (see the introduction of \citealt{PetreHwang2007} and references therein for a thorough list),  enough that 0540's pulsar is commonly referred to as a `twin' of the Crab. However, the Crab lacks a visible shell in any wavelength \citep{Hester2008}, whereas 0540 has a prominent shell with a diameter of about 55\arcsec\ in radio \citep{ManchesterStaveley-Smith1993} and X-ray. The X-ray shell was first detected with the \textit{ROSAT} HRI by \citet{SewardHarnden1994}. The first spatially-resolved X-ray spectroscopic studies of the shell with the \textit{XMM-Newton} RGS \citep{van-der-HeydenPaerels2001} and with the \textit{Chandra} ACIS \citep{HwangPetre2001} revealed that the X-ray shell is characterized by cool ($kT=0.5-1.5$ keV) thermal material in the west.  \citeauthor{HwangPetre2001} also found much hotter thermal gas ($kT \simeq 4$ keV) in the eastern half of 0540, as well as nonthermal arcs in the east and west of the SNR. Both of these studies found elemental abundances roughly consistent with LMC values throughout the shell, suggesting that the SNR displays little X-ray emission from ejecta and has already undergone substantial interaction with surrounding ISM. 

The ejecta surrounding the PWN have been studied as well, mostly in optical. A bright ring of O-rich material about 3\arcsec\ thick surrounds the PWN in optical [\ion{O}{3}] observations \citep{MathewsonDopita1980,KirshnerMorse1989,MorseSmith2006}, but this ring is not apparent in the X-ray. \citet{MorseSmith2006} also detected a diffuse halo of [\ion{O}{3}] emission around the PWN, out to a radius of 8\arcsec\ ($\sim 2.2$ pc) from the pulsar. Long-slit spectroscopy of this halo showed features that \citeauthor{MorseSmith2006} interpreted as heavy velocity broadening of the [\ion{O}{3}] line, consistent with a velocity dispersion of 3300 km s$^{-1}$. \citeauthor{van-der-HeydenPaerels2001} reported a tentative detection of O-rich material, specifically the \ion{O}{8} Ly$\alpha$ line at 656 eV, in the direction of the PWN. The authors did not provide quantitative information on the strength of the line, but the detection of O-rich material emitting in the X-ray has not been confirmed by subsequent X-ray studies.  \citet{WilliamsBorkowski2008} analyzed \textit{Spitzer} IRAC, MIPS, and IRS observations of 0540 and found that the emission from the pulsar and PWN shows an excess of flux above what would be expected from the IR synchrotron spectrum from the pulsar and PWN. The authors interpreted this flux excess as emission from warm dust within the PWN that has condensed out of ejecta.

\citet{PetreHwang2007} analyzed a 34 ks \textit{Chandra} ACIS observation to examine the spatial variation of X-ray emission within the PWN itself. These authors found that the photon index of the power-law emission from the PWN varies with increasing radius, from $\Gamma = 1.4$ immediately surrounding the pulsar to $\Gamma = 2.4$ at the edge of the PWN, at a radius of 1.2 pc (5\arcsec). This softening of the spectrum with increasing radius, which is due to synchrotron losses by the highest energy particles as they flow out to the edge of the PWN, is seen in other young PWNe as well, such as the Crab \citep{WillingaleAschenbach2001}, where the photon index varies from 1.8 near the pulsar to 2.4 at a radius of 1.2 pc. \citeauthor{PetreHwang2007} also found that X-ray spectra from outside a 5\arcsec\ radius required a combination of power-law and thermal plasma models to adequately fit the observed emission.

The most comprehensive survey of the shell to date was by \citet{ParkHughes2010}, who used a 118 ks \textit{Chandra} ACIS observation to examine several regions in the shell of 0540, mainly large areas in the dim south and east portions of the shell. The authors found conclusive evidence of synchrotron radiation from hard ``arc" features at the extreme east and west of the SNR shell. They also analyzed regions in the east, south, and west of the shell, and found conditions mostly consistent with those found by earlier studies: hot thermal or non-thermal material in the east and cooler shocked thermal material in the west, all of which show no sign of ejecta. However, \citeauthor{ParkHughes2010} report marginal evidence for the existence of Fe-rich material in the south, which these authors interpreted as a clump of ejecta that is located near the southern edge of the blast wave.
 
Although extensive theoretical modeling of the early evolution of PWNe and their interaction with the expanding SNR shell has been done (e.g. \citealt{BlondinChevalier2001,van-der-SwaluwAchterberg2001,van-der-SwaluwDownes2004,BucciantiniAmato2004,GelfandSlane2009}) there are surprisingly few young PWNe that are suitable for observing these effects \citep{Chevalier2005}, due either to the lack of a visible shell, as in the Crab, or due to a low galactic latitude, which restricts the range of observable wavelengths. 0540 is nearby enough to resolve spatial detail, located at a high galactic latitude (and thus relatively unabsorbed), and has a prominent shell. A comparison between two young PWNe such as the Crab and 0540 opens several interesting lines of inquiry. What does the presence of circumstellar material around 0540 say about its likely evolutionary path? How does 0540's environment influence the morphology and spectral properties as a function of wavelength? Does the active star-forming environment around the SNR affect the evolution of the shell? Are environmental effects in the shell mirrored in the PWN? Is a comparison between 0540 and the Crab reasonable?

In this study, we use the high spatial resolution data from \textit{Chandra} to identify and analyze regions in the shell that have not been previously studied. We also detail the plasma conditions within the noticeable X-ray halo surrounding the PWN and interior to the shell for the first time. This analysis increases our understanding of interactions within the shell, while probing the possibility for interactions between the PWN and the surrounding ejecta.

\section{Observations}
 \subsection{X-ray Observations}\label{sec:Xray_obs}
Three observations of 0540, with a total integration time of 118 ks, were made with the \textit{Chandra} Advanced CCD Imaging Spectrometer (ACIS) on 15 February 2006, 16 February 2006, and 18 February 2006. This data set was used to analyze the western and southern portions of the shell in \citet{ParkHughes2010}. These observations were archived as ObsID 5549, 7270, and 7271, respectively. In each of these observations, the entire SNR was imaged on a $1/4-$subarray of the ACIS-S3 back-illuminated chip, which is particularly sensitive to soft ($<$2 keV) X-rays. The use of a $1/4-$subarray was done in order to mitigate the effects of pileup from the bright central pulsar by reducing the frame time from the standard ACIS value of 3.2 s to approximately 1.0 s \citep{ParkHughes2010}. These observations achieved an unbinned spatial resolution $\lesssim 1$\arcsec\ and a spectral resolution ($\Delta E/E$) of $\sim 5-20$ over the main energy band for the ACIS-S detectors ($0.3-8.0$ keV).The earlier 28 ks ACIS observation of 0540 \citep{HwangPetre2001,PetreHwang2007} was not included in our survey. This observation was taken in the ACIS FAINT mode (instead of the VFAINT mode used in the three 2006 observations) in order to study the central PWN. However, these data did not provide any noticeable improvement in photon statistics in the shell, so we did not include them in our study.

\begin{figure}
\epsscale{0.75}
% \plotone{f1.eps}
\plotone{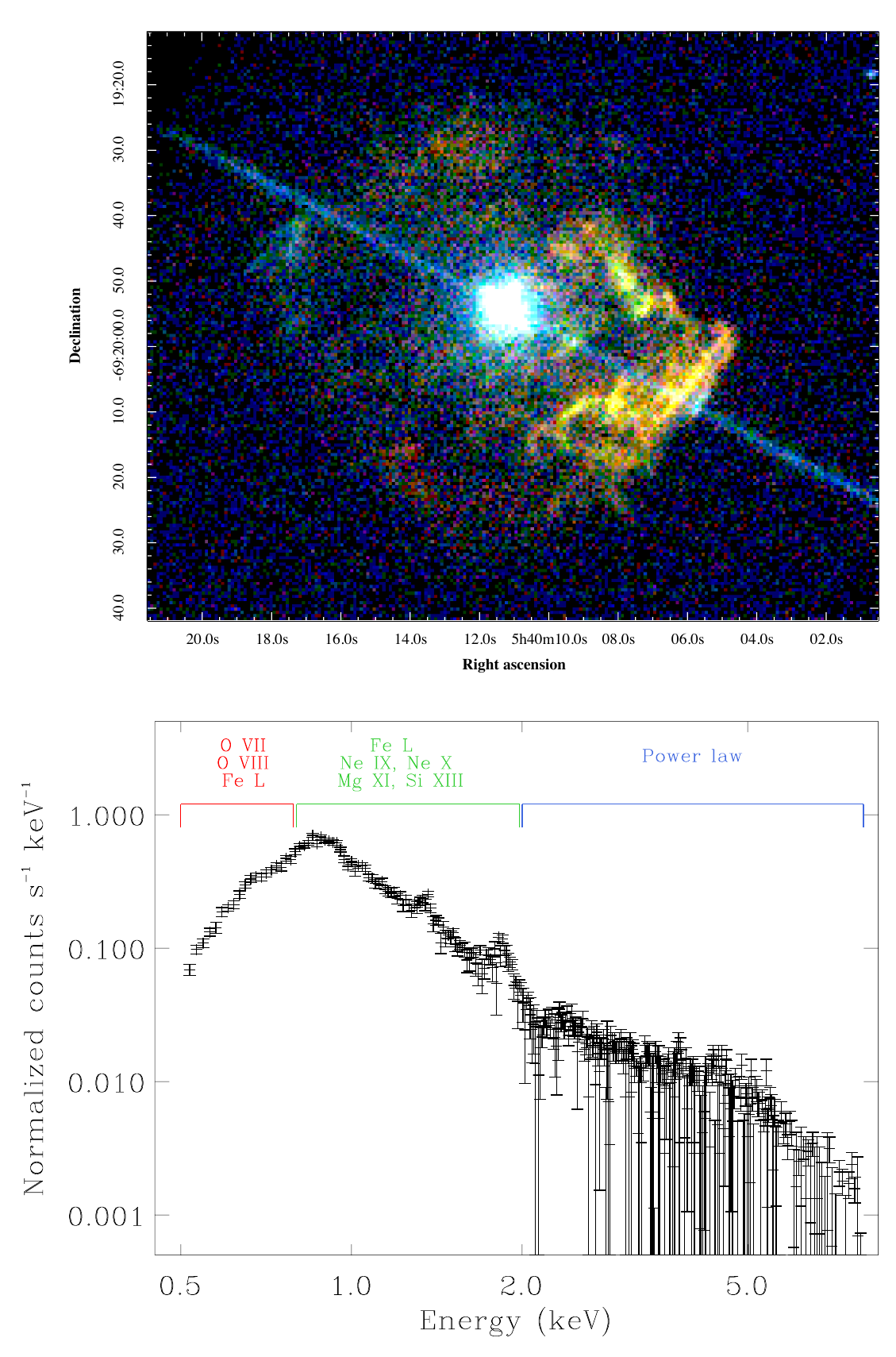}
\caption{Top: False color X-ray image of SNR 0540-69.3, split into energy bands of 0.3-0.8 keV (red), 0.8-2.0 keV (green), and $>2.0$ keV (blue). The image has been smoothed using a Gaussian with a radius of 3 pixels. The intensity values are scaled using a square-root scale from 0-6 counts per pixel (red and blue) and 0-20 counts per pixel (green). Bottom: Total spectrum of the shell ($>5$\arcsec\ from the pulsar) with energy bands and prominent lines within those bands highlighted.}
\label{fig:CXO_color}
\end{figure}

The level 1 event files were reprocessed using the \texttt{chandra\_repro} tool in CIAO 4.4 \citep{CIAO2006}, with calibration information from CALDB 4.5.1. This tool automatically creates new level 2 event files, accounting for charge transfer inefficiency, good time intervals, and time-dependent gain variations. These level 2 event files were used for all analysis.  A false-color image, created by splitting the X-ray emission into soft ($0.3-0.8$ keV), medium ($0.8-2.0$ keV), and hard ($>2.0$ keV) bands, is shown in Figure \ref{fig:CXO_color}. This image was produced using the CIAO tool \texttt{reproject\_events} to reproject the three \textit{Chandra} observations to a common plane and co-add them. The energy bands are chosen to highlight spectral windows dominated by specific emission lines or processes evident in the spectrum of the entire remnant, also shown in Figure \ref{fig:CXO_color}. We have chosen the soft band to cover the lowest-energy X-ray lines, principally H and He-like oxygen (\ion{O}{7} at 574 eV and the \ion{O}{8} doublet at 653 and 654 eV), although lines in the iron L-shell such as \ion{Fe}{17} (725, 727, 739 eV) and \ion{Fe}{18} (771 eV) are prominent in this band at low temperatures ($kT \sim 0.5-1.0$ keV). Fe L shell lines dominate in the medium band, but highly ionized states of silicon (\ion{Si}{13} at 1865 eV; prominent above $kT = 1.0$ keV) and magnesium (\ion{Mg}{11} at 1352 eV), are present. Neon (\ion{Ne}{9} at 922 eV and \ion{Ne}{10} at 1022 eV) also appears, although \ion{Ne}{9} is only prominent at low temperatures. It is apparent from even a cursory glance at Figure \ref{fig:CXO_color} that these lines are minor components in the spectrum, either due to a lack of emitting ejecta or to a low ionization state, and that continuum emission is the dominant component in the spectrum. We expect that if a nonthermal  continuum is present, it will be most apparent in the hard band. 

The X-ray data are complicated somewhat by the fact that the pulsar and the central $\sim 3$\arcsec\ of the PWN are heavily affected by pileup, despite the use of a subarray to reduce the ACIS frame time. A related effect is the ACIS readout streak, which can be seen prominently in Figure \ref{fig:CXO_color}. The streak is caused by photons from a bright source being collected while the CCD is reading out, which smears out the source along its chip rows. To prevent the displaced photons from contaminating spectra in the interior, we use the CIAO tool \texttt{acisreadcorr} to block out the contaminated rows. 

\subsection{Radio Observations}\label{sec:Radio_obs}
We use radio-continuum observations from Australian Telescope Compact Array (ATCA) project C014. Details of the observations can be found in Table \ref{tbl:ATCA_observations}. The radio galaxy PKS $1934-638$ was used as a primary flux calibration source for all observations, with the radio sources PKS $0454-810$ and PKS $0530-727$ used for phase calibration. The \textsc{miriad}\footnote{\url{http://www.atnf.csiro.au/computing/software/miriad}} \citep{SaultTeuben1995} and \textsc{karma} \citep{Gooch1995} software packages were used for reduction and analysis. Images were formed using \textsc{miriad} multi-frequency synthesis \citep{SaultWieringa1994} and natural weighting. They were deconvolved with primary beam correction applied. The same procedure was used for both \textit{U} and \textit{Q} Stokes parameter maps. Figure \ref{fig:ATCA_combined} shows surface brightness maps for the ATCA data.

\begin{figure}
\epsscale{1.0}
% \plotone{f3.eps}
\plotone{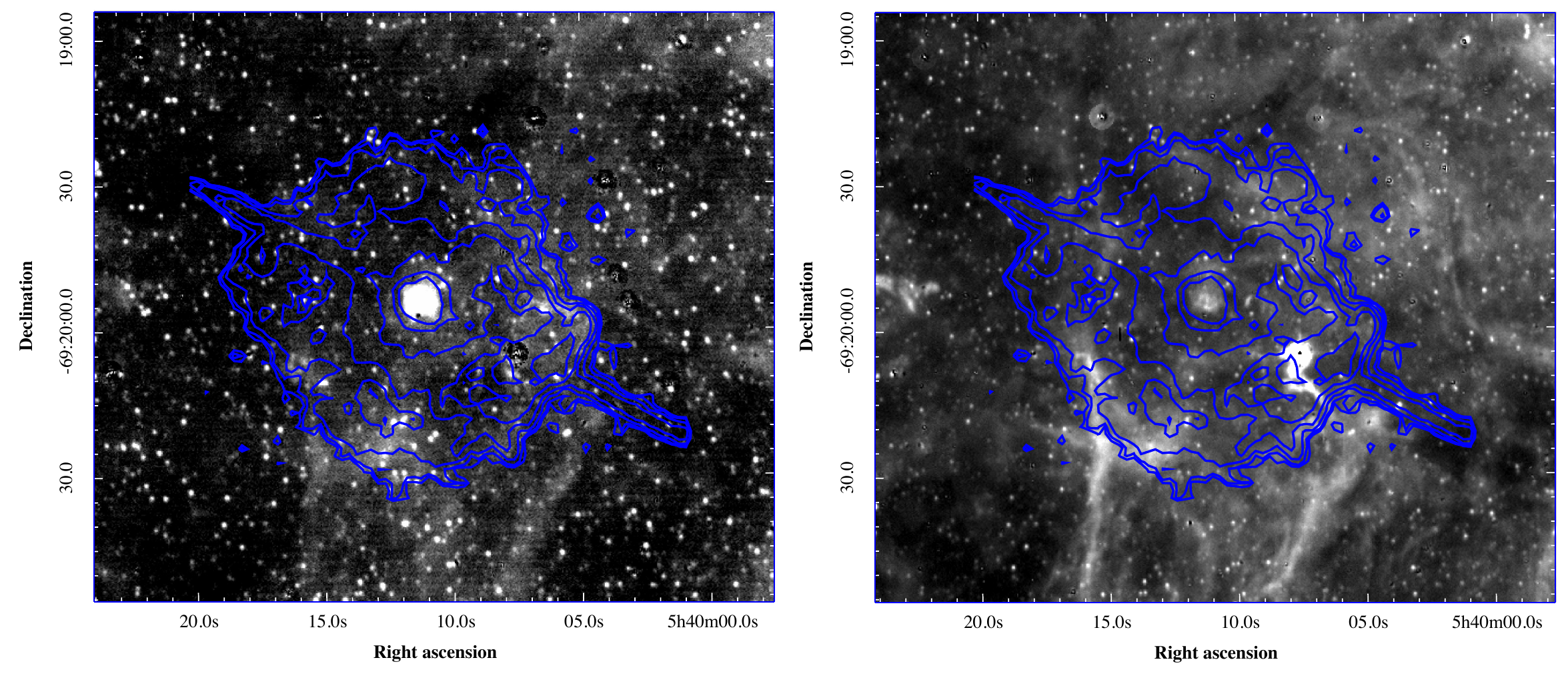}
\caption{Left: ESO image of SNR 0540-69.3 in [\ion{O}{3}], with bright field stars removed via PSF subtraction. Right: ESO image of SNR 0540-69.3 in H$\alpha$, with bright field stars removed via PSF subtraction. Some residuals from subtracted stars are still visible in both images. X-ray contours are overlaid in blue; the linear features extending to the northeast and southwest in the contours are caused by the readout streak. The bright features to the south and southwest is the northern edge of the superbubble surrounding the OB association LH 104 \citep{LuckeHodge1970}.}%, visible in Figure \ref{fig:CXO_color}. The bright feature to the south is }
\label{fig:optical}
\end{figure}

Optical images of 0540 were obtained from the European Southern Observatory (ESO) 30 Doradus/WFI Data Release 1.0. These images were taken with the Wide Field Imager on the 2.2 m telescope at the ESO's La Silla Observatory. For the main band used, [\ion{O}{3}], 5 exposures with a total exposure time of 1500 s were aligned, sky-subtracted, and co-added by the VOS/ADP team, providing a FWHM of 0.82\arcsec. For the H$\alpha$ band, 4 exposures with a total exposure time of 4800 s were combined, providing a FWHM of 0.73\arcsec. Further details on the filters, data reduction and processing may be found at \url{http://archive.eso.org/archive/adp/ADP/30_Doradus/}. We removed bright field stars from the [\ion{O}{3}] and H$\alpha$ images in order to clarify the faint nebular structure via PSF subtraction, using standard procedures with the tasks \texttt{daofind}, \texttt{phot}, \texttt{psf}, and \texttt{allstar} in the IRAF DAOPHOT package. The PSF-subtracted [\ion{O}{3}] and H$\alpha$ images can be seen in Figure \ref{fig:optical}.

\subsection{IR Observations}\label{sec:IR_obs}
Infrared images of 0540 were obtained from the \textit{Spitzer} Heritage Archive. These images were taken with the \textit{Spitzer} Infrared Array Camera (IRAC) as part of observing program ID 3680. The observations resulted in a 26.8 s integration time in all four IRAC bandpasses (3.5 $\mu$m, 4.5 $\mu$m, 5.8 $\mu$m, and 8 $\mu$m) with a pixel size of $\sim$1.2\arcsec. These observations were retrieved after undergoing standard pipeline processing and are presented here without additional reprocessing beyond minor cropping and rescaling. The shortest-wavelength IR images reveal little of interest and are mostly dominated by field stars. In 5.4 and 8 $\mu$m, however, more features become apparent, as shown in Figure \ref{fig:Spitzer_tricolor}, which shows a false-color image of 0540 from the IRAC data, with the X-ray contours overlaid.

\section{Morphology}\label{sec:morphology}

\subsection{X-ray Morphology}
As shown in Figure \ref{fig:CXO_color}, the X-ray morphology shows a patchy and incomplete shell at a radius of about 30\arcsec\ from the pulsar, with the bright central PWN surrounded by a ``halo" of X-ray emission out to a radius of 9\arcsec. This image reveals that the character of the emission varies greatly depending on position. The western half of the SNR is dominated by softer emission with a more obvious shell morphology. The shell is substantially brighter in the southwest than in the northeast. The eastern half consists mostly of diffuse harder emission. The hard and nonthermal ``arcs" investigated by \citet{ParkHughes2010} are also apparent at the extreme eastern and western boundaries of the SNR shell. The PWN is apparent as a bright and heavily piled-up feature at the center of the shell, elongated slightly towards the NW. The elongation was initially thought \citep{GotthelfWang2000} to be a jet, but high-resolution optical imaging \citep{MorseSmith2006} showed a diffuse ``breakout" instead of a collimated jet in this region. Other analysis \citep{SandinLundqvist2013} has suggested that this elongation may actually be a torus around the pulsar, similar to that seen around the Crab \citep{WeisskopfHester2000}, with the jet pointing to the southwest. 

\subsection{Radio Morphology}
\begin{figure}
\epsscale{0.75}
% \plotone{f2.eps}
\plotone{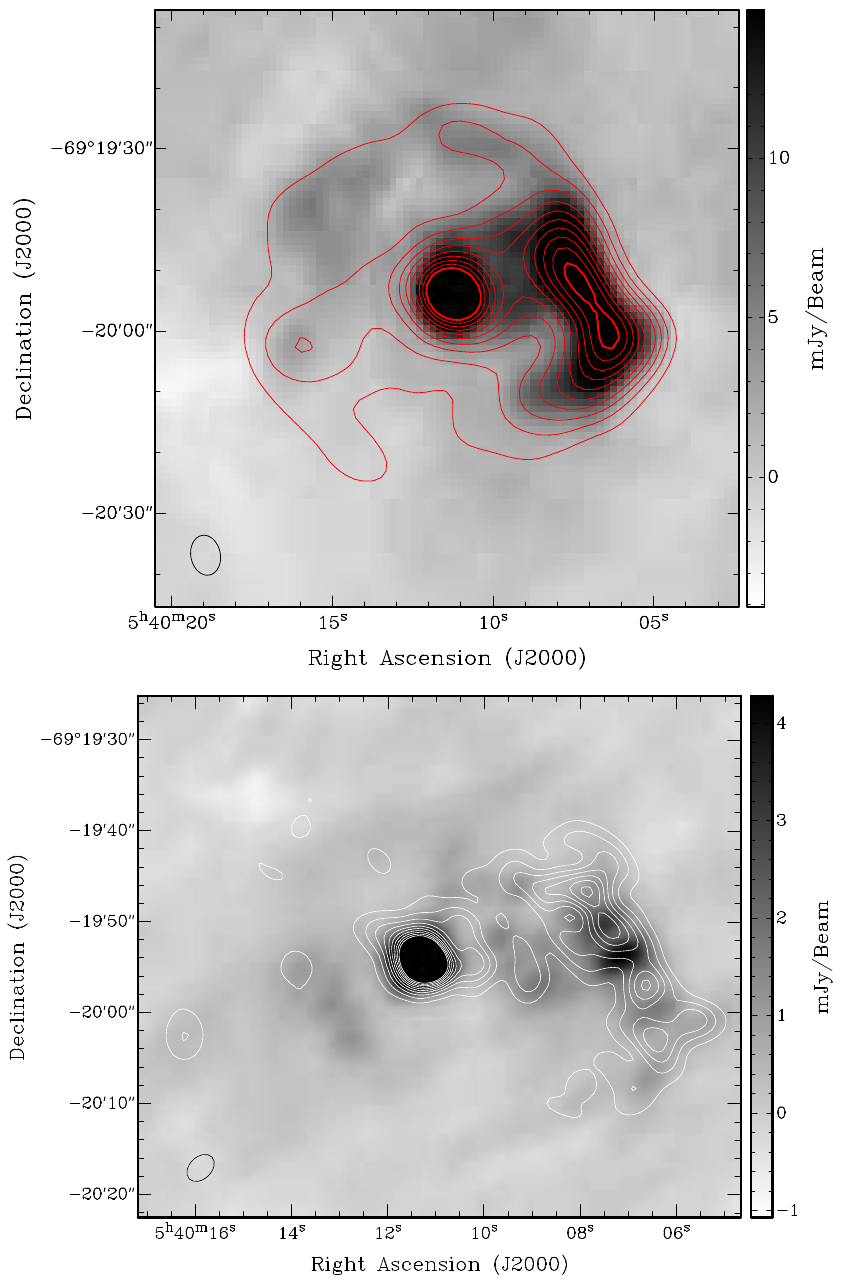}
\caption{Top: ATCA observations of SNR 0540-69.3 at 13 cm (2290 MHz) in grayscale, overlaid with 20 cm (1513 MHz) contours in red. Bottom: ATCA observations of SNR 0540-69.3 at 3 cm (8895 MHz) in grayscale, overlaid with 6 cm (5824 MHz) contours in white. Both images show contours from 3$\sigma$ to 30$\sigma$ in steps of 3$\sigma$. The black ellipse in the lower left represents the synthesized beamwidth of 6.6\arcsec\ $\times$ 4.9\arcsec\ at 13 cm and 3.5\arcsec\ $\times$ 2.4\arcsec\ at 3 cm. The sidebar quantifies the pixel map in units of mJy beam$^{-1}$.}
\label{fig:ATCA_combined}
\end{figure}

The visible radio structure is strongly dependent on frequency, as shown by the maps in Figure \ref{fig:ATCA_combined}. The 1513 MHz (20 cm) radio map shows a nearly circular area of diffuse radio emission about 1\arcmin\ in diameter, surrounding the extremely bright ($>100$ mJy beam$^{-1}$) PWN. This diffuse emission shows a mostly constant intensity around 8-10 mJy beam$^{-1}$. The western edge of the shell appears to be somewhat flattened, although given the large beam size at this frequency (9.7\arcsec$\times$7.9\arcsec), determining morphology on such a small scale is difficult. This edge shows a concentration of considerably brighter emission ($\sim 60$ mJy beam$^{-1}$) about 20\arcsec\ long. The 2290 MHz (13 cm) radio map shows a distinct shell morphology in the northern portion of the shell. The eastern and northern portions of the shell ($5-6$ mJy beam$^{-1}$) are much fainter than the western portion of the shell ($>15$ mJy beam$^{-1}$), which appears much as it does in the 1513 MHz image. The 5824 MHz (6 cm) image also reveals that the western edge of the SNR appears to be composed of a number of bright ($5-10$ mJy beam$^{-1}$) knots. At the highest frequency radio band surveyed, 8895 MHz (3 cm), the shell is mostly below the background noise threshold, and only the PWN and the knots along the western edge of the shell are visible. We find possible structure of the PWN in the 5824 MHz image (Figure \ref{fig:ATCA_combined}), where there appears to be elongation of the emission in the NE-SW direction.

\subsection{Optical Morphology}
\begin{figure}
  % \plotone{f5.eps}
\plotone{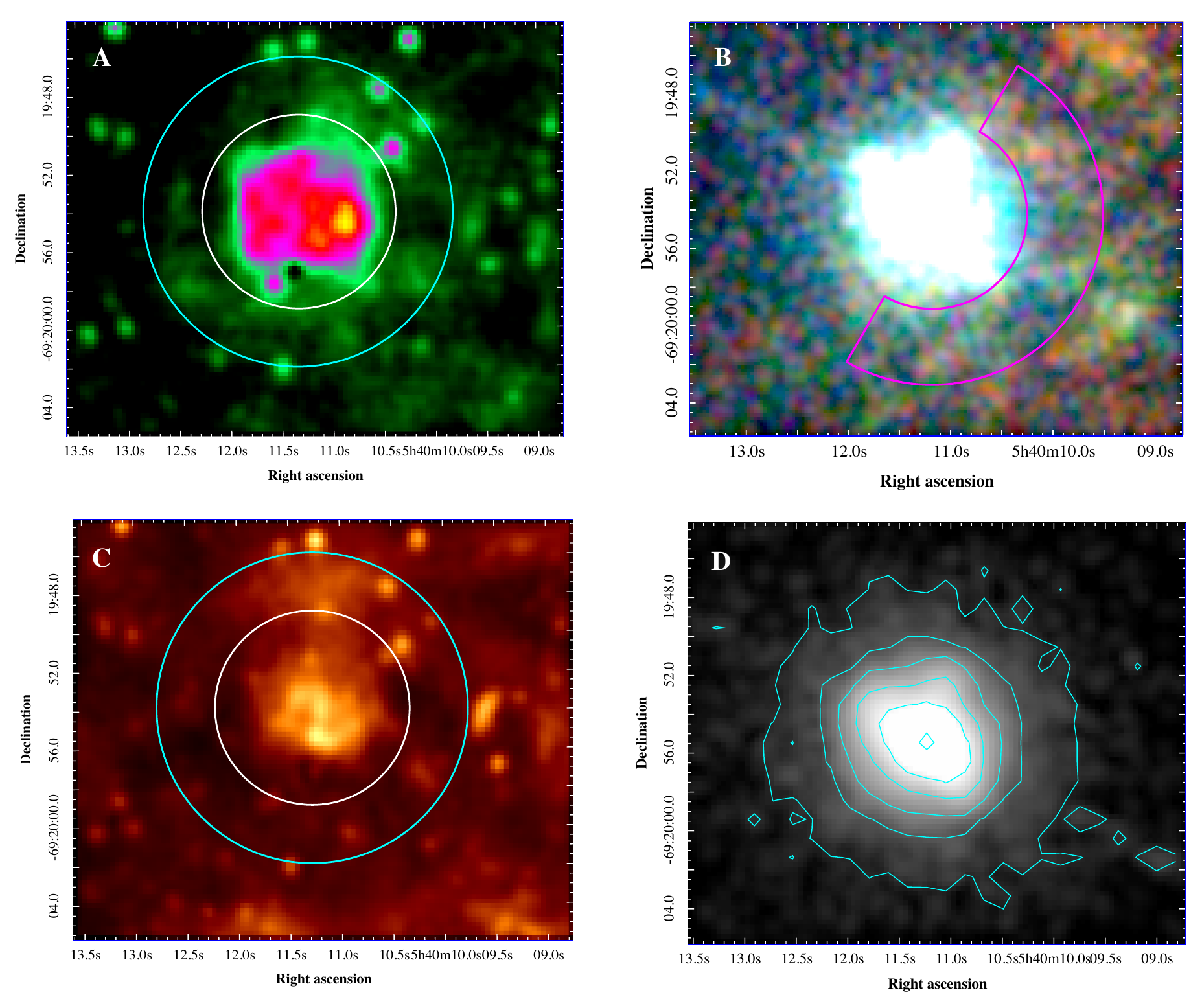}
\caption{A: ESO detail of the halo region in [\ion{O}{3}]. The white circle is drawn at a radius of 5\arcsec, the approximate extent of the X-ray PWN. The blue circle is drawn at a radius of 8\arcsec, the approximate extent of the extended emission noted by \citet{MorseSmith2006}. B: \textit{Chandra} detail of the halo region in RGB X-ray, with the same bands as Figure \ref{fig:CXO_color}. C: ESO detail of the halo region in H$\alpha$. The white and blue circles overlaid on the image are as in the [\ion{O}{3}] image. D: Detail of the halo region from the simulated MARX observation. Contours are drawn at levels of 0.6, 3, 13, 54, and 217 counts per pixel.}
\label{fig:tiny_color}
\end{figure}

At optical wavelengths, the field is very complicated. Data from CO surveys \citep{OttWong2008,WongHughes2011} indicates that 0540 is located in a large ridge of molecular clouds to the east of 30 Doradus, and the \ion{H}{2} region N158, containing the OB association LH 104, is located less than 2\arcmin\ to the south \citep{DunnePoints2001}. In both H$\alpha$ and [\ion{O}{3}], shell structure is difficult to distinguish from this complex surrounding environment. However, diffuse [\ion{O}{3}] emission appears to roughly trace the SNR shell in the north and south, at a flux level\footnote{Flux levels are quoted in terms of ADU. One ADU (analog to digital unit) corresponds to one count recorded by the CCD detector.} barely above the background ($0.03-0.05$ ADU s$^{-1}$, with a background flux of $0.02-0.03$ ADU s$^{-1}$). The circular pattern becomes lost in diffuse emission at the east and west.  A detail of the halo region in H$\alpha$ and [\ion{O}{3}] is shown in Figure \ref{fig:tiny_color}. The bright 8\arcsec\ ring around the PWN is clearly visible ($\sim 0.24$ ADU s$^{-1}$) in the [\ion{O}{3}] images, as is the fainter ($0.06-0.07$ ADU s$^{-1}$) extended emission out to about 5\arcsec\ seen with HST long-slit spectroscopy by \citet{MorseSmith2006}. The latter feature is not visible in the HST WFPC2 images. West of the PWN in the shell region, the diffuse [\ion{O}{3}] emission noted by \citet{MathewsonDopita1980} is visible. 

\subsection{IR Morphology}\label{sec:IR_morphology}
\begin{figure}
  % \plotone{f4.eps}
\plotone{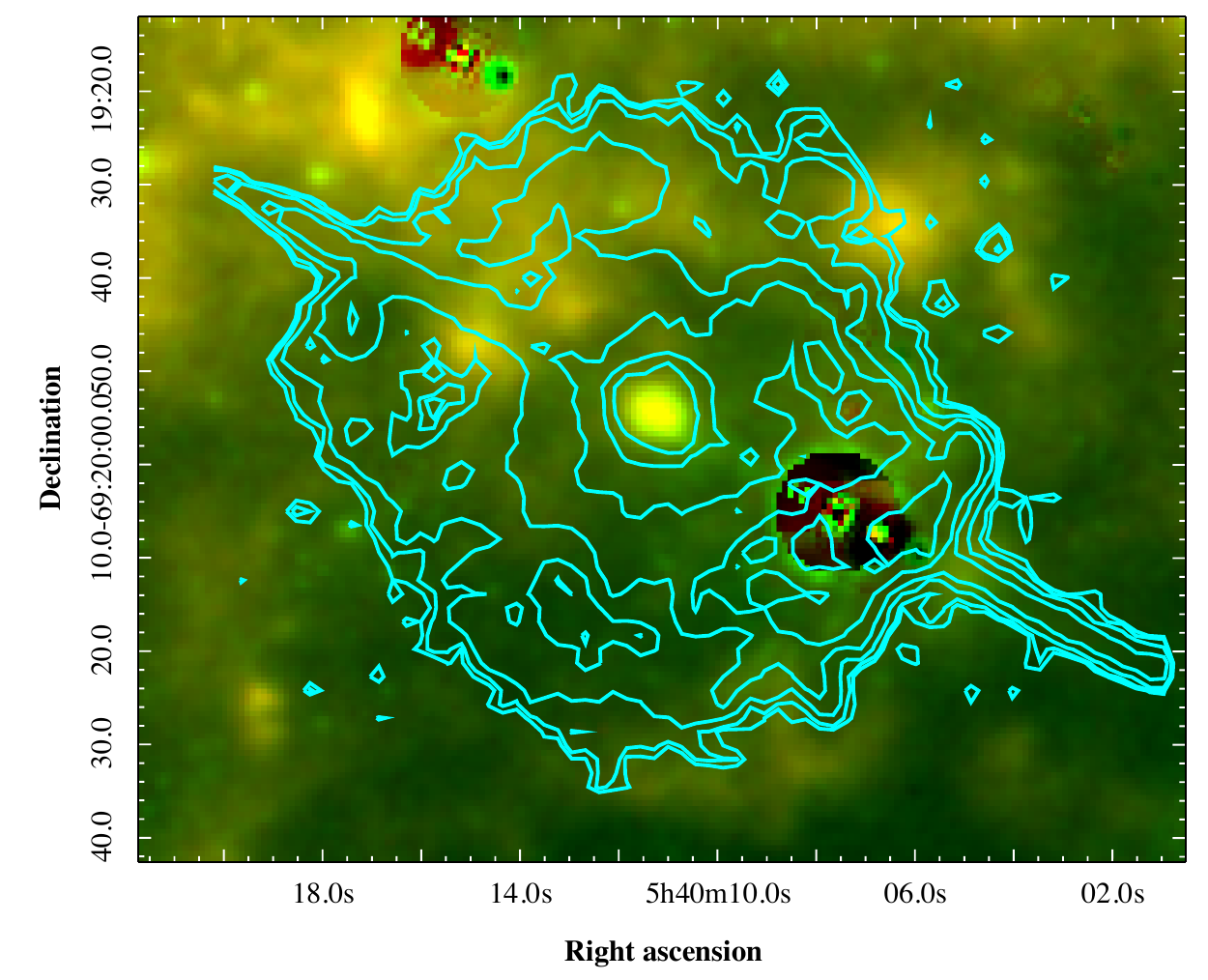}
\caption{False-color image of Spitzer IRAC data. Red is 8.5 $\mu$m IRAC data scaled linearly from 3.5-10.8 mJy/sr and green is 5.6$\mu$m IRAC data scaled linearly from 0.1-4.4 mJy sr$^{-1}$. X-ray contours are overlaid in blue. As in Figure \ref{fig:optical}, the long linear features extending northeast and southwest from the shell are caused by the readout streak in the X-ray data. Some residuals are visible from PSF subtraction of bright field stars.}
\label{fig:Spitzer_tricolor}
\end{figure}

The unrelated LMC star 2MASS J05400761-6920049 is prominent in the west of the shell in IR, making fine detail in this region difficult to see. (This star has been removed via PSF subtraction in Figure \ref{fig:Spitzer_tricolor}.) Bright diffuse emission in the northeast is visible, along with two starlike objects. Neither the diffuse emission or starlike objects appear to be correlated to the SNR shell morphology in radio and X-ray. North of this, when combined with the [\ion{O}{3}] emission, a portion of the shell becomes apparent. A circular structure which appears suggestive of the SNR shell is also visible outside the east limb and in the south. However, \citet{WilliamsBorkowski2008} state that this apparent shell structure in the IR is spectroscopically indistinguishable from the background and argue, based on this, that its relation to the SNR may be coincidental. However, Figure \ref{fig:Spitzer_tricolor} shows a portion in the southwest of the shell that closely follows, and is just exterior to, the X-ray contours. This suggests that the IR structure may be a portion of the shell radiating mostly in IR, perhaps caused by interaction with a local molecular cloud. We use this possible association with caution. 

For both the optical and IR data, definitively determining whether specific features are associated with the SNR is difficult.  Due to these considerations, we only make use of these data in a qualitative manner, to lend further insight to the quantitative conclusions drawn from the radio and X-ray data in the following sections.

\section{Results}
We detail the analysis results of our multi-wavelength data campaign here and in the following sections. Due to the large number of results, we will present them in the following order. First, we give the spectral index, polarization, and rotation measure values calculated from our radio observations, in Section \ref{sec:radio_spec_index}. We then use these results to motivate a comparison between radio and X-ray morphology in Section \ref{sec:radio_vs_xray}. We also use the spectral index calculation to motivate a discussion of the broadband spectral energy distribution of 0540 in Section \ref{sec:broadband_spec}. In Sections \ref{sec:specextract} and \ref{sec:xray_fitting}, we use detailed X-ray spectral analysis to elucidate the correlations brought forth in the previous sections. In Section \ref{sec:discussion}, we discuss the various wavelengths together to form a coherent picture of 0540.

\subsection{Radio Results}\label{sec:radio_spec_index}
Because our radio observations cover several different frequencies, we may calculate the spectral index $\alpha$, where the flux density $S \propto \nu^\alpha$, from the radio flux measurements. 
We measured the flux density of the core of 0540 at four wavelengths (20 cm, 13 cm, 6 cm, and 3 cm), using the observations listed in Table \ref{tbl:ATCA_observations}. However, as interferometers such as ATCA lack the zero-spacing measurement responsible for the large-scale emission, we were unable to directly measure the extended emission from the SNR shell. These observations were particularly unsuited for measuring extended emission due to the long baselines used, where the shortest distance between any two antennas ranged from 77 m to 337 m. As an alternative, we measured the integrated flux density for the entire SNR from archival radio continuum observations. We estimate the 36 cm measurements from the Molonglo Synthesis Telescope (MOST) mosaic image (as described in \citealt{MillsTurtle1984}) and the 36 cm Sydney University Molonglo Sky Survey (SUMMS) image \citep{MauchMurphy2008}. The 20 cm flux density was measured from a mosaic image published by \citet{HughesStaveley-Smith2007} while the 6 cm and 3 cm flux densities were measured from mosaics published by \citet{DickelMcIntyre2010}.

\begin{figure}
  % \plotone{f6.eps}
\plotone{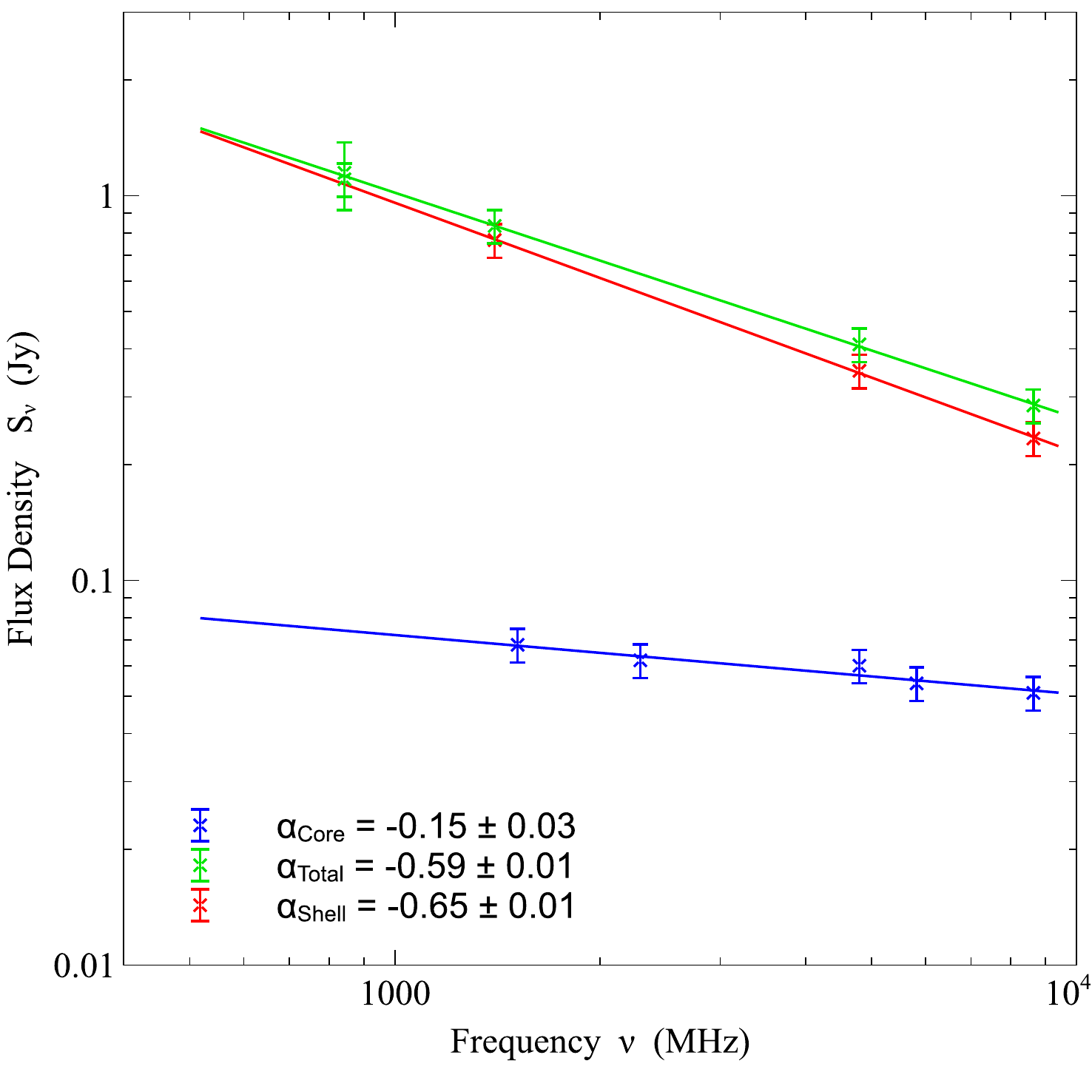}
\caption{Radio spectral index measurements for the PWN ($\alpha_{\text{Core}}$, shown in blue), the shell ($\alpha_{\text{Shell}}$, shown in red), and the entire radio source ($\alpha_{\text{Total}}$, shown in green).}
\label{fig:wide_spec_index}
\end{figure}

 As we lacked high resolution images with short baselines, we deduce the flux density of the remnant's shell by subtracting the value of the core from the measurement for the entire SNR. The resulting values for the entire core and shell are shown in Table \ref{tbl:radio_flux} and then used to produce the spectral index plot in Figure \ref{fig:wide_spec_index}. Errors in these measurements predominantly arose from defining the edge of the PWN. Using the flux densities in Table \ref{tbl:radio_flux}, we estimate a spectral index for the PWN of $\alpha_{Core} = -0.15 \pm 0.03$. This is consistent with, but much more refined than, the earlier estimate of the spectral index by \citet{ManchesterStaveley-Smith1993} of $\alpha = -0.25 \pm 0.10$. Our value is also consistent with the typical radio spectral index for a PWN of $\alpha \geq -0.3$ and with spectral index values of other known PWNe such as G$326.3-1.8$ ($\alpha = -0.18$; \citealt{DickelMilne2000}) and IKT 16 ($\alpha = -0.19 \pm 0.2$; \citealt{OwenFilipovic2011}). We also estimate for the entire SNR of $\alpha_{Total} = -0.59 \pm 0.01$, and a spectral index for the shell alone of $\alpha_{Shell} = -0.65 \pm 0.01$. Our value for the shell is consistent with the typical value \citep{Green2009} for synchrotron radiation from an SNR.

\begin{figure}
  % \plotone{f7.eps}
\plotone{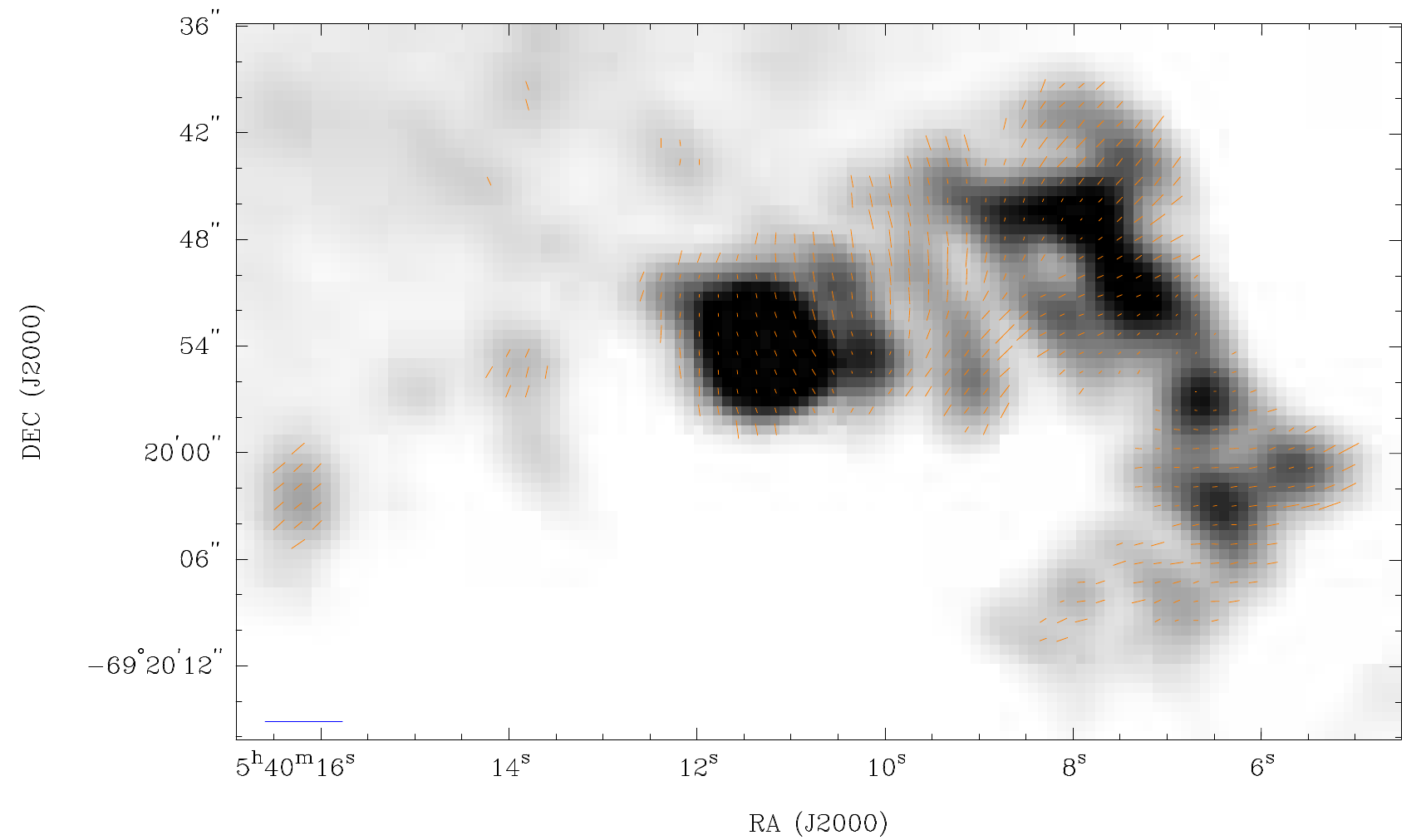}
\caption{Electric field vectors of SNR 0540-69.3 at 6 cm (5824 MHz). The line in the lower left-hand corner represents a polarization of 100\%.}
\label{fig:E_field}
\end{figure}

The fractional polarization, $P$, was calculated at 6 cm, using the equation
\begin{equation}
P = \frac{\sqrt{S_Q^2 + S_U^2}}{S_I}
\end{equation}
where $S_Q$, $S_U$, and $S_I$ are the integrated intensities for the $Q$, $U$, and $I$ Stokes parameters. We estimate that the mean fractional polarization of the shell at $P=11 \pm 3$\%. The resulting electric field vectors for the 5824 MHz observations are shown in Figure \ref{fig:E_field}.

\begin{figure}
  % \plotone{f8.eps}
\plotone{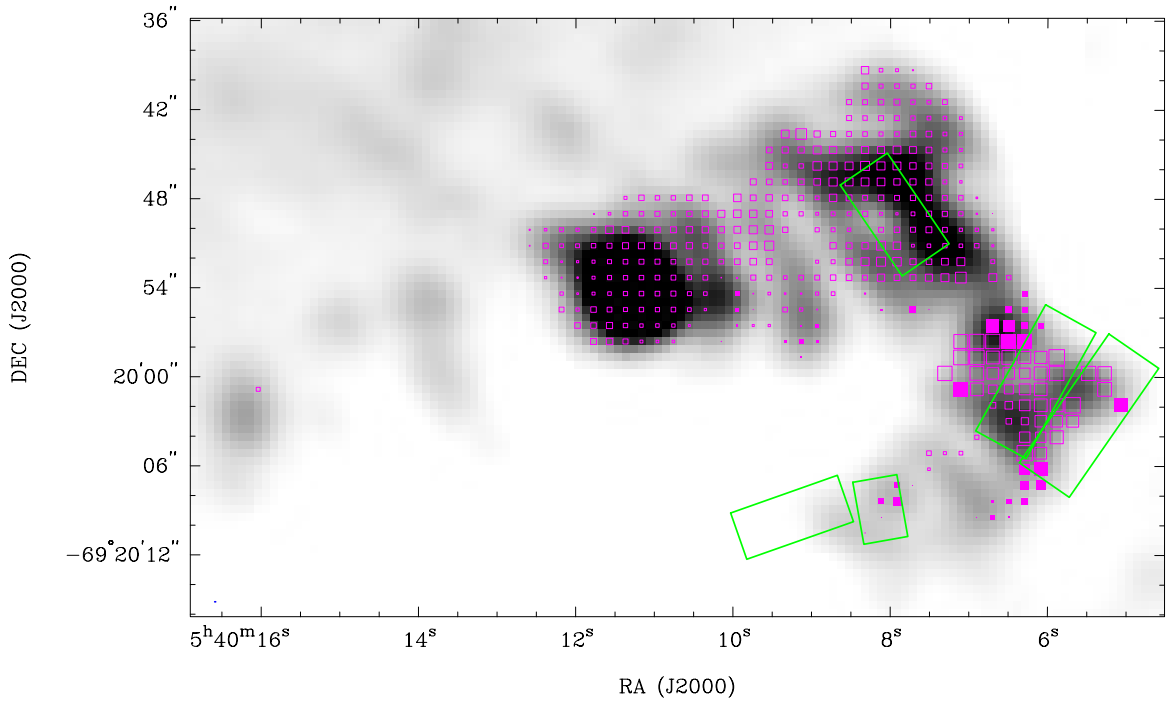}
\caption{Faraday rotation of SNR 0540-69.3 between 5824 MHz and 4786 MHz overlaid on the 5824 MHz intensity image. Filled purple boxes represent positive rotation measure, while empty purple boxes represent negative rotation measure. A box dimension of 0.5\arcsec\ represents a rotation measure of 394 rad m$^{-2}$. Regions for X-ray spectral extraction are overlaid in green.}
\label{fig:FaradayRM}
\end{figure}

\begin{figure}
  % \plotone{f9.eps}
\plotone{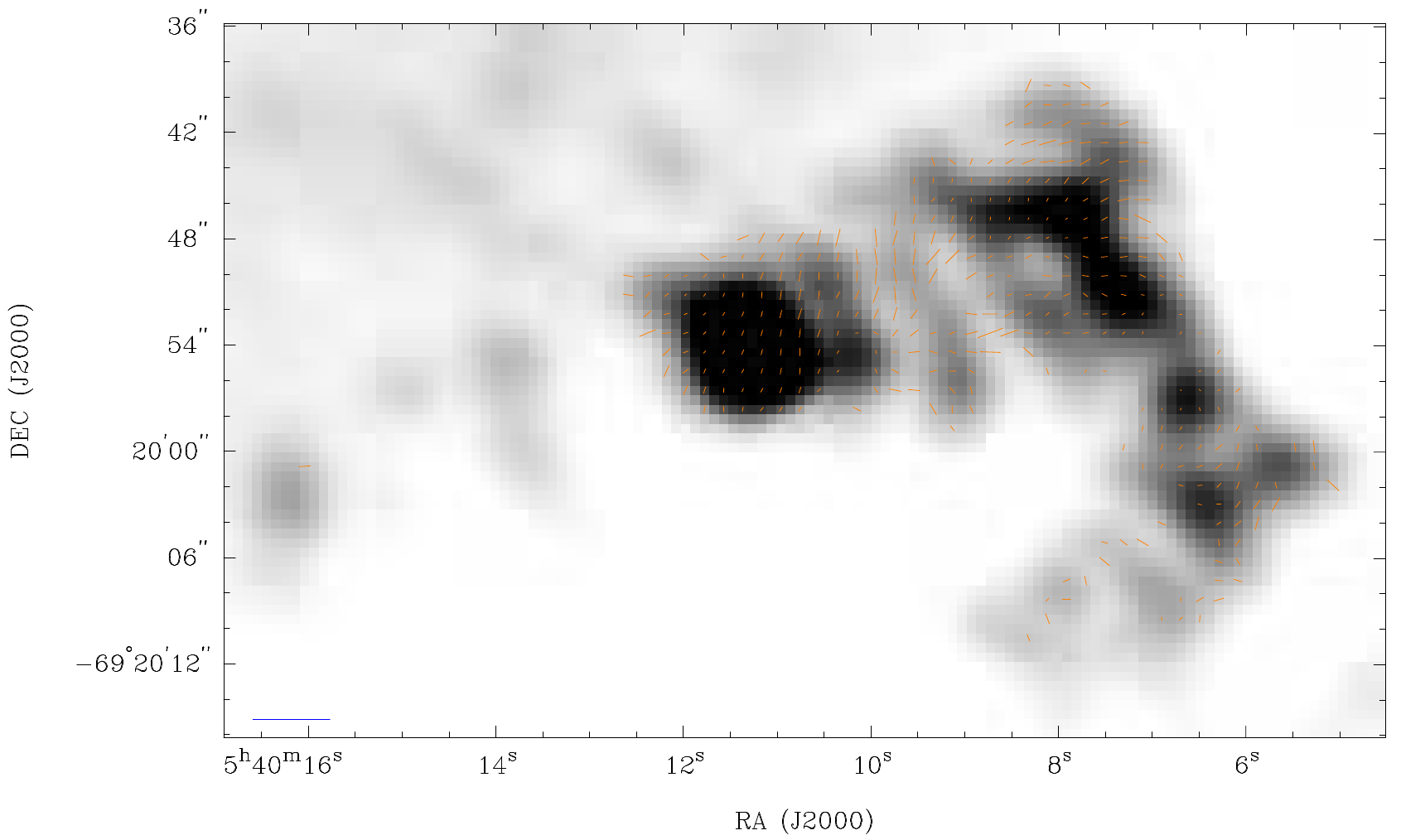}
\caption{Magnetic field vectors of SNR 0540-69.3. The vector length represents the polarized intensity at 6 cm (5824 MHz). The line in the lower left-hand corner represents a polarization of 100\%.}
\label{fig:B_field}
\end{figure}

Polarization position angles were taken from both 6 cm observations (5824 and 4786 MHz) and used to estimate the value of the Faraday rotation measure. The resulting rotation measures are plotted in Figure \ref{fig:FaradayRM}, with the open boxes representing negative values of rotation measure and the filled-in boxes representing positive rotation measure. The average rotation measure for the central plerion was $-346$ rad m$^{-2}$, while the shell averaged a slightly higher value of negative rotation, $-369$ rad m$^{-2}$. The rotation measure was then used to find the zero-wavelength, or intrinsic, values by de-rotating the measured position angles. The vectors in Figure \ref{fig:B_field} show the intrinsic values and directions of the magnetic field in the remnant.

\subsection{Comparison of X-ray and Radio Emission Morphology}\label{sec:radio_vs_xray}
\begin{figure}
  % \plotone{f10.eps}
\plotone{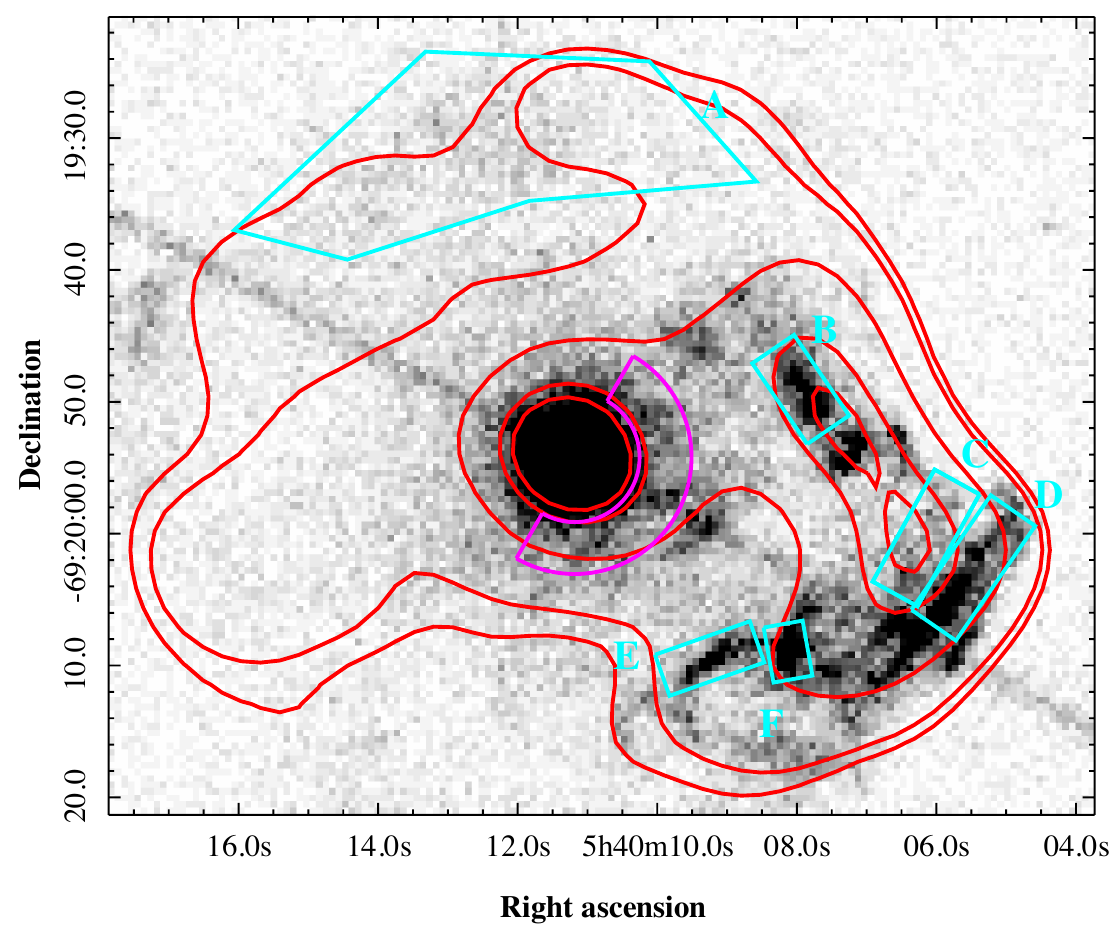}
\caption{\textit{Chandra} grayscale image of SNR 0540-69.3, unbinned and scaled linearly from 0-25 counts per bin. 1513 MHz ATCA contours are overlaid in red at levels of 7.8, 10, 26, 52, and 63 mJy beam$^{-1}$. Regions for X-ray spectral extraction are overlaid in cyan.}
\label{fig:CXO_radio}
\end{figure}

Figure \ref{fig:CXO_radio} shows the X-ray emission in grayscale, with radio contours at 1.5 GHz and 5 GHz overlaid. Two areas of coincident emission are immediately apparent: the PWN and the western edge of the shell. The apparent size of the PWN in radio and X-ray is about the same: 5\arcsec\ across in radio and X-ray, although determining the exact size in radio is difficult, as the radio beam size in our observations is comparable. The equivalent radio and X-ray sizes are at odds with the tendency of most PWNe, such as the Crab \citep{Hester2008} to decrease in apparent size when seen at increasing energy. A possible cause of this is an unusually low magnetic field, such as that seen in 3C 58 \citep{GreenScheuer1992,SlaneHelfand2004}, which has a magnetic field of $\sim 80$ $\mu$G. As discussed below in Section \ref{sec:broadband_spec}, we see some evidence of a magnetic field this weak in 0540.

The western edge of the shell is another area of coincidence between X-ray and radio emission. In both X-ray and radio, the shell is sharply flattened and brightened on this edge. The knot about 20\arcsec\ northeast of the PWN is the most 
prominent brightened area in 5 GHz radio and in X-ray. However, in 1.5 GHz radio, the western edge features two teardrop-shaped knots that have no obvious counterpart in X-ray or in higher-frequency radio observations. This suggests that two different populations of electrons are responsible for the dissimilar areas of emission. 

There are a few areas of emission where the radio and X-ray emission do not coincide at all. The noticeable area of radio emission in the southeast portion of the shell, visible as a knot in the 5 GHz map and as an amorphous extension of the shell in 1.5 GHz, has no obvious counterpart in X-ray. The extended area of emission in the northern area of the shell in the 1.5 GHz map also has no counterpart in 5 GHz or in the X-ray. Conversely, the synchrotron-emitting X-ray knot in the east \citep{ParkHughes2010} has no obvious counterpart in any surveyed radio band. One possibility, in this eastern knot, is that the radio synchrotron emission is blocked by the Razin effect \citep{GinzburgSyrovatskii1965,RybickiLightman1979}, where synchrotron emission below a cutoff frequency $\nu_R$ is suppressed by thermal plasma in the emitting region. The cutoff frequency is given by \citep{Melrose1980,Dulk1985} $\nu_R = 20n_e/B$, where $n_e$ is the electron density in cm$^{-3}$ and $B$ is the magnetic field in Gauss. 
The highest radio frequency surveyed, 8895 MHz, shows no evidence of this synchrotron knot, so we can use this as a lower limit of the potential Razin cutoff frequency. The magnetic field in this knot is $B=20-140$ $\mu$G \citep{ParkHughes2010}, which means that suppressing synchrotron emission below 8895 MHz would require a minimum electron density in the emitting region of $\sim 0.01 - 0.06$ cm$^{-3}$. The minimum density found in the east of 0540 by \citeauthor{ParkHughes2010} is $\sim 0.17$ cm$^{-3}$, so this is not an unreasonable requirement. 

\subsection{PWN Broadband Spectrum and Magnetic Field}\label{sec:broadband_spec}
\begin{figure}
  % \plotone{f11.eps}
\plotone{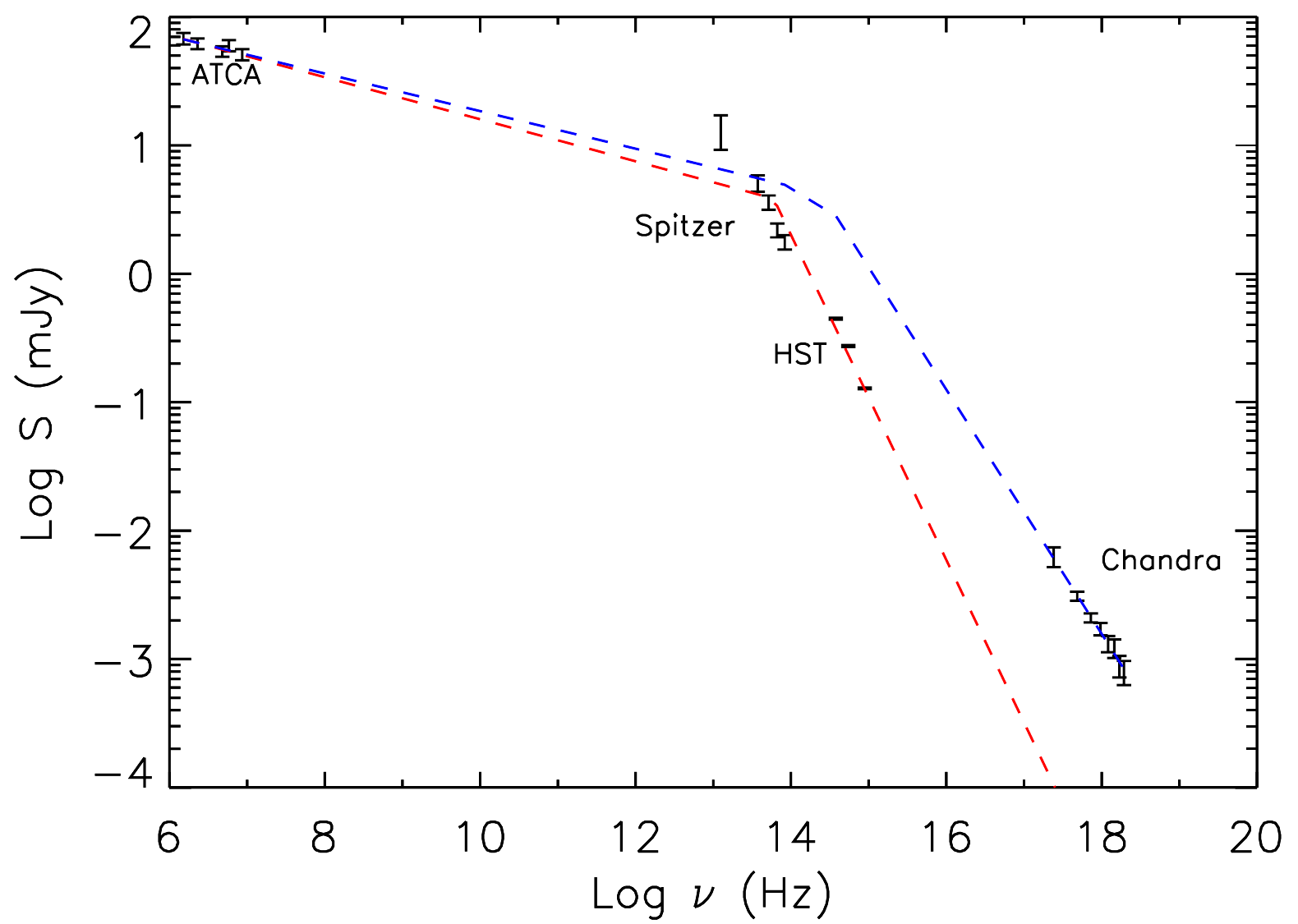}
\caption{Broadband spectrum of the PWN and pulsar of SNR 0540-69.3. Radio data is from this work. IR flux measurements are from \citet{WilliamsBorkowski2008}, optical flux measurements are from \citet{SerafimovichShibanov2004}, and X-ray flux measurements are from \citet{KaaretMarshall2001}. The best-fit broken power law to radio/IR/optical is plotted in red, with a break frequency of $6 \pm 1 \times 10^{13}$ Hz and spectral indices $\alpha_{lo} = -0.16 \pm 0.1$ and $\alpha_{hi} = -1.27 \pm 0.3$. The best-fit broken power law to the radio and X-ray data, with a break frequency of $2.5^{+56}_{-2.3} \times 10^{14}$ Hz and spectral indices $\alpha_{lo} = -0.14^{+0.1}_{-0.08}$ and $\alpha_{hi} = -0.95 \pm 0.13$, is plotted in blue. Note the flux excess at about $10^{13}$ Hz, attributed by \citet{WilliamsBorkowski2008} to emission from warm dust in the PWN.}
\label{fig:broadband}
\end{figure}

The wide frequency range covered by past and present observations of 0540 makes it possible for us to examine the broadband spectrum of the PWN. This spectrum, covering the range $10^6 - 10^{19}$ Hz, is shown in Figure \ref{fig:broadband}. It is apparent from Figure \ref{fig:broadband} that although individual wavebands may be accurately fit with a power-law, as evidenced by the radio spectral index calculation in Section \ref{sec:radio_spec_index}, a single power-law does not accurately describe the entire broadband spectrum. The break in the spectrum between the radio and IR bands is attributed \citep{GinzburgSyrovatskii1965} to synchrotron losses from higher energy electrons that have exceeded their synchrotron lifetime. 

The synchrotron lifetime for an electron radiating at a frequency $\nu$ can be approximated \citep{Pacholczyk1970} as $\tau_s = 6 \times 10^{11} B^{-3/2} \nu_b^{-1/2}$ where $\tau_s$ is the synchrotron lifetime in seconds, $B$ is the PWN magnetic field in Gauss, and $\nu$ is the frequency in Hertz. At the break frequency, the synchrotron lifetime is equal to the age of the PWN, so we can determine the PWN magnetic field from a determination of the break frequency. 
Fitting this broadband spectrum to a broken power-law (plotted in Figure \ref{fig:broadband}) results in a break frequency of $6 \pm 1 \times 10^{13}$ Hz. Assuming a nebular age \citep{Reynolds1985} of 1100 years, this results in a magnetic field strength of $250 \pm 15$ $\mu$G. This is consistent with the magnetic field estimate of \citet{ManchesterStaveley-Smith1993}, who estimated a magnetic field strength of 250 $\mu$G for an age of 1100 years, and substantially lower than that of \citet{PetreHwang2007}, who estimated a PWN equipartition magnetic field of $740$ $\mu$G based on an RXTE measurement of the pulsar braking index together with models of PWN evolution from \citet{Reynolds1985}. 

\citet{SerafimovichShibanov2004} have pointed out that there appears to be a second break, or ``knee" in the spectrum of the pulsar between optical and X-ray wavelengths. This effect, which appears in a few other pulsars such as Vela \citep{ShibanovKoptsevich2003} and Geminga \citep{MignaniCaraveo1998}, may be caused by depressed optical emission, perhaps due to unresolved cyclotron absorption lines \citep{MignaniCaraveo1998} and propagation effects on the outflow of relativistic particles from the pulsar. \citet{WilliamsBorkowski2008} also noted the presence of dust inside the PWN, so the depressed optical flux may also be caused by reddening from dust particles within the PWN itself. \citeauthor{SerafimovichShibanov2004} also find a knee in the spectrum of the PWN after subtracting the pulsar contribution. We therefore attempted another broken power-law fit to the radio and X-ray data alone. This fit, plotted in blue in Figure \ref{fig:broadband}, results in a break frequency of $2.5^{+56}_{-2.4} \times 10^{14}$ Hz, which in turn gives a PWN  magnetic field of $106^{+171}_{-68}$ $\mu$G. Although poorly constrained, this raises the possibility that the similar size of the PWN from radio through X-ray could be caused by a very weak magnetic field, leading to correspondingly longer synchrotron lifetimes.

\section{X-ray Spectral Extraction and Modeling}\label{sec:specextract}
\begin{figure}
\epsscale{0.75}
% \plotone{f12.eps}
\plotone{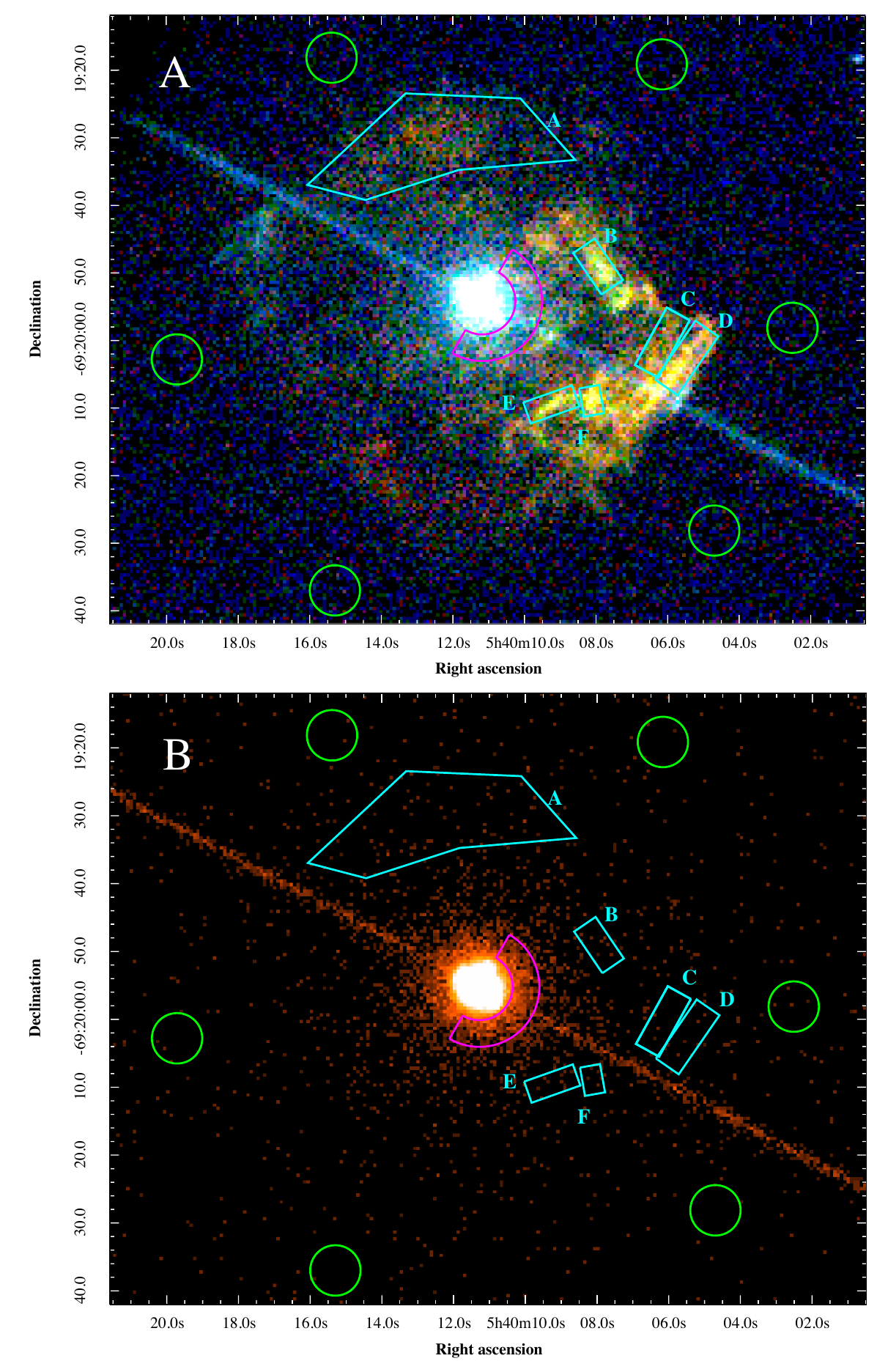}
\caption{A: False-color X-ray image as shown in Figure \ref{fig:CXO_color}, with extraction and background regions overlaid. B: Simulated MARX observation of SNR 0540-69.3, with extraction and background regions overlaid.}
\label{fig:marx_sim}
\end{figure}

Thus far we have covered conditions in 0540 in a largely qualitative manner. In this and the following section, we use detailed X-ray spectral analysis to provide quantitative information on the plasma conditions in the areas discussed above.
The X-ray spectrum of 0540 outside of the pulsar and PWN cannot be described by a pure power-law. Analysis of the central PWN in \citet{PetreHwang2007} showed that a thermal component becomes prominent in the X-ray spectrum at a radius of 5\arcsec\ from the pulsar. The combination of thermal and power-law spectra raises the possibility of interaction between the PWN and surrounding material becoming apparent at this 5\arcsec\ radius. In addition to investigating the presence of thermal plasma in a $5-9$\arcsec\ radial region, we perform azimuthal clockings of a half-annular region to determine if the dichotomous morphology in the shell is mirrored in the halo. We extract identical regions rotated every 30 degrees counter-clockwise. The region oriented at our nominal 0 degrees (aligned with the readout streak) is overlaid on the \textit{Chandra} RGB image in Figure \ref{fig:tiny_color}, with a wider angle view shown in Figure \ref{fig:marx_sim}.

Spectra were also extracted from regions in the SNR shell, shown in blue in Figure \ref{fig:marx_sim}. These regions were chosen to probe previously unexplored regions in the SNR shell and, together with the results of \citet{ParkHughes2010}, \citet{HwangPetre2001}, and \citet{PetreHwang2007}, to complete the picture of 0540's interactions with the surrounding environment. Region A covers a large, diffuse, and fairly soft feature in the extreme northeast edge of the shell. Region B covers a small bright region in the NW edge of the shell. This region is similar to the NW region analyzed by \citet{ParkHughes2010}. However, those authors included the point source CXOJ$054007.2-691954$ \citep{EvansPrimini2010} in their extracted region. This source appears starlike in all four IRAC bands, as well as in B and V band optical images, and we therefore conclude that it is an unrelated object. Region B is chosen to exclude this source, in order to provide a clearer picture of the shell emission in the NW. Region D covers a linear bright feature that marks the western edge of the shell in X-ray and radio, while Region C examines the area immediately interior to Region D. Finally, Regions E and F probe an interesting feature in the southwest, with Region E covering a ``hot spot" of relatively hard and bright emission, and Region F covering a nearby bright shock-like feature. For all shell and interior regions, the background spectra were extracted collectively from the group of six small circular regions around the edge of the SNR shell (shown in green in Figure \ref{fig:marx_sim}) to provide an average background. Although different groups of background regions, as well as blank-sky backgrounds, were tried, the choice of background regions did not have any noticeable effect on the resultant spectral fits.

Spectra for each extraction region were extracted from the reprocesssed level 2 event files using the \texttt{specextract} tool in CIAO 4.4. This tool automatically extracts source and background spectra and creates appropriate redistribution matrix and ancillary response files for each region. The source and background spectra were extracted separately for each observation and then added together using the CIAO tool \texttt{combine\_spectra}. The combined spectra were binned to a minimum of 20 counts per bin to allow Gaussian statistics to be used for error analysis. Data below 0.5 keV and above 8 keV are discarded, owing to calibration uncertainties in the ACIS detectors.

We use the CIAO spectral modeling package \textit{Sherpa} to fit the extracted spectra. The shell regions should be a one- or two-temperature thermal plasma with or without a synchrotron component \citep{ParkHughes2010}. For emitting material that has been shocked recently or is comparatively low-density, we expect the material to be in non-equilibrium ionization (NEI). We use the \textit{xsvnei} model in \textit{Sherpa}\footnote{Models beginning with the prefix \textit{xs} in \textit{Sherpa} are implementations of the corresponding model in XSPEC. For instance, the \textit{xsvnei} model in \textit{Sherpa} implements the \textit{vnei} model in XSPEC.} \citep{BSB1994,BLR2001,HSC1983,LOG1995} to model these conditions. We modify this model slightly by using an updated line list developed by Kazik Borkowski that includes additional detail for inner shell processes, particularly Fe-L lines. Details of this line list may be found in \citet{BadenesBorkowski2006}. We also test the \textit{xsvpshock} plane-parallel shock model, which allows for a range of different densities and shock ages in the emitting material. If, on the other hand, the emitting material has been shocked a long time ago or is comparatively high-density, we expect the material to be in collisional ionization equilibrium (CIE). We model these conditions with the \textit{xsvapec} model, which relies on the AtomDB 2.0.2 code \citep{SBLR2001, FosterJi2012} to model the emission spectrum. The extracted spectra are dominated by thermal emission from $0.5-2$ keV, with a higher energy ``tail" in Regions A, B, and D that dominates emission above 2 keV. Emission from thermal plasmas at temperatures of $0.5-2$ keV, typical of the thermal emission from 0540's shell \citep{HwangPetre2001}, are dominated by lines of O, Ne, Mg, Si, and Fe at these energies. All fits begin with all elemental abundances frozen to typical LMC ISM values\footnote{These values, relative to solar, are He=0.89, C=0.26, N=0.16, O=0.32, Ne=0.42, Mg=0.74, Si=1.7, S=0.27, Ar=0.49, Ca=0.33, Fe=0.5, and Ni=0.62.} \citep{RD1992}. Elemental abundances are allowed to vary from LMC ambient values if changing the abundances significantly improves the fit, as shown via $F$-test  (a more complex model is allowed if the null hypothesis probability $P<0.05$). To fit the harder ``tail" extending to $\sim 5$ keV in Regions A, B, and D, we test a variety of additional fit components: a simple power-law model (\textit{xspowerlaw} in \textit{Sherpa}), a cutoff synchrotron spectrum (\textit{xssrcut} in \textit{Sherpa}; \citealt{Reynolds1998,RK1999}), and an additional high-temperature thermal component.

As demonstrated by \citet{SerafimovichShibanov2004}, the use of a single absorption component to account for the absorbing column along the line of sight to 0540 is incorrect. Although the bulk of the absorption takes place locally in the LMC, there is a small but significant amount of galactic absorption along this line of sight as well. We thus convolve our spectral models with a two-component photoelectric absorption model with variable elemental abundances (\textit{xsvphabs}). One component is fixed at solar abundances \citet{AG1989} and frozen to the galactic \ion{H}{1} column density in the direction of 0540, $0.07 \times 10^{22}$ atoms cm$^{-2}$ \citep{DL1990}. The second component has abundances fixed to values typical of the LMC ISM \citep{RD1992}. We allow the absorbing column density to vary freely for this component when fitting regions in the shell, although we reject fits with an LMC column density greater than $0.8 \times 10^{22}$ atoms cm$^{-2}$ or lower than $0.3 \times 10^{22}$ atoms cm$^{-2}$, the upper and lower limits, respectively, of the \ion{H}{1} column density in the LMC \citep{KimStaveley-Smith2003}. When fitting the interior regions around the PWN, allowing the LMC column density to vary freely frequently results in it being driven to unphysically low or high values, so for these we fix the LMC column at the approximate value of LMC absorption along the line of sight to the pulsar, $n_H=0.5 \times 10^{22}$ atoms cm$^{-2}$ \citep{SerafimovichShibanov2004}. For all models, we use the dielectronic recombination rates of \citet{MMCV1998} with cross-sections from \citet{BCMC1992}.

Extracting spectra is complicated by contamination from the pulsar/PWN PSF, which is quite large due to the brightness of the pulsar and PWN. The \textit{Chandra} PSF is highly energy-dependent and becomes much broader at higher energies\footnote{A thorough discussion of the PSF can be found in the \textit{Chandra} Proposer's Observatory Guide (\url{http://asc.harvard.edu/proposer/POG/html/ACIS.html})}. To account for this, we simulated observations of the pulsar and PWN using MARX 5.0.0. These simulations used a source with a surface brightness distribution based on the spectral index and flux measurements from the HRC observation of 0540 \citep{KaaretMarshall2001}. An image of the simulated observation may be seen in Figure \ref{fig:marx_sim}. For each analyzed region, a spectrum was extracted from the corresponding region in the simulated observation to examine the level of contamination in that region. Apart from Region A, the contamination from the pulsar and PWN PSF is negligible in the shell regions and was thus ignored. For Region A in the shell, as well as the annular and wedge regions within the halo, the contamination from the PWN and pulsar PSF makes a significant contribution to the observed spectrum. We accounted for this by fitting the spectrum extracted from the simulated observation to an unabsorbed power-law, which was then included as a model component when fitting the observed spectrum from the corresponding region, with the power-law index and the normalization fixed to the best fit values. As the overall number of counts for these contamination spectra was low, we fit the data unbinned, using the C-statistic in \textit{Sherpa} \citep{Cash1979}. For region A, the contamination spectrum was fit by a power-law with a photon index $\Gamma = 0.11 \pm 0.3$, normalized to $3.5^{+1.3}_{-1.0} \times 10^{-7}$ photons keV$^{-1}$ cm$^{-2}$ s$^{-1}$ at 1 keV. For the wedge regions, the contamination spectrum was fit by a power-law with $\Gamma = 1.0$, normalized to $9.9 \times 10^{-6}$ photons keV$^{-1}$ cm$^{-2}$ s$^{-1}$ at 1 keV. Errors quoted are 90\% confidence intervals for these fits.

\section{X-ray Spectral Fitting}\label{sec:xray_fitting}

\subsection{Shell Regions}
\begin{figure}
%  \plotone{f13.eps}
\plotone{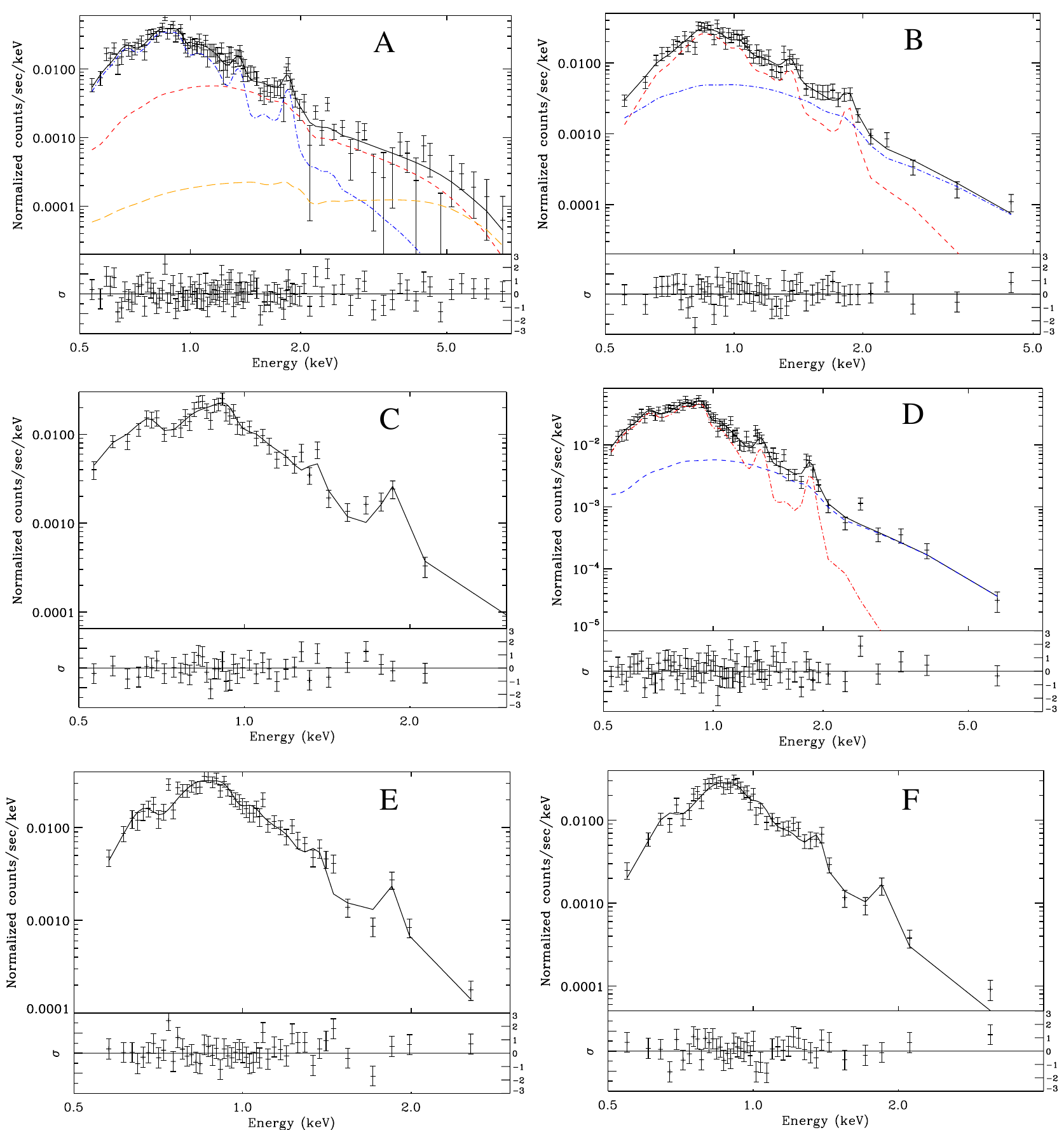}  
\caption{Spectra with best-fit models and fit residuals for Regions A-F in the shell. For regions with multiple fit components, the power-law component is shown in blue and the thermal component is shown in red. The contamination from the pulsar's PSF is shown in orange in the plot for region A.}
\label{fig:shell_spectra}
\end{figure}

The best-fit models for each shell region are shown in Table \ref{tbl:nei_fits}. Errors are 90\% (1.6$\sigma$) confidence intervals, calculated using the \texttt{conf} tool in \textit{Sherpa}.  This tool allows all thawed parameters to vary while calculating the confidence interval to find the most general error bounds. We find metal abundances consistent with those of the LMC ISM, although we find slight evidence for depletion in Mg and Si in nearly every region, and a slight enhancement in Ne in Region D. We also find that the surveyed regions are fit by a single-temperature NEI model, with a harder component evident in regions at the edge (A, B, and D). For every shell region surveyed, the equilibrium \textit{vapec} model resulted in a substantially worse ($\chi^2_{red} = 1.5-2$ or more) fit than an NEI model, required unrealistically high or low values for electron temperature or column density, or both. Extracted spectra for these regions, with best fit models overlaid, are shown in Figure \ref{fig:shell_spectra}. We first discuss the lower temperature thermal component of each shell region individually and then discuss the fits to the higher temperature components in regions A, B, and D.

Region A covers a large enough area that contamination from the PWN PSF contributes a significant amount of flux, which is accounted for by a fixed power-law, shown in orange in part A of Figure \ref{fig:shell_spectra}. This region is chosen to cover the dim northwest portion of the shell, which has not been studied in detail by previous \textit{Chandra} studies of 0540. The lower-temperature thermal component is best fit by an NEI plasma with an electron temperature of $kT = 0.6^{+0.2}_{-0.1}$ keV. Single-component thermal models did not result in an acceptable fit ($\chi^2_r \gtrsim 2$), and purely nonthermal models provided a similarly poor fit. The best-fit thermal component requires somewhat depleted abundances of Mg ($\sim 4\sigma$ depletion) and Si ($3 \sigma$ depletion) relative to typical LMC values.

Region B covers a small bright feature along the NW edge of the shell. As discussed earlier, this region is similar to the NW region analyzed by \citet{ParkHughes2010}, with the exception of a point source included in the extraction region by \citeauthor{ParkHughes2010} As stated in Section \ref{sec:specextract}, based on the starlike appearance of this source in IR through X-ray wavebands, we believe this source to be an unrelated object. The spectrum of the point source is hard, with emission extending out to $\sim 5$ keV, which may have led \citeauthor{ParkHughes2010} to the conclusion that the spectrum of this region is harder than it actually is.  The best fit for the thermal component in Region B is obtained with an NEI plasma with an electron temperature of $kT=0.60^{+0.04}_{-0.06}$ keV and slightly depleted Mg ($1\sigma$ depletion) and Si ($4 \sigma$ depletion), similar to conditions seen in Region A. The spectrum of the NW region examined by \citeauthor{ParkHughes2010} is fit by those authors with a two-temperature NEI model, but the harder of the two temperature components in their fit requires an extremely low ionization timescale ($n_e t < 8 \times 10^8$ cm$^{-3}$ s) which is hard to reconcile with the observed density in this region, as discussed in Section \ref{sec:shell_density} below, and with the age of 0540.

Region C The spectrum in Region C drops off sharply at 2 keV. As in the other shell regions, the best-fit thermal model is an NEI plasma, but the electron temperature is about 2$\sigma$ higher ($kT = 1.1 \pm 0.3$ keV) than in other shell regions. The unusually high temperature offers the possibility that this region may have two distinct thermal components, but none of the two-temperature models we investigated resulted in an improved fit. A lower-temperature thermal model plus a power-law or other nonthermal model resulted in unrealistically steep values for the powerlaw photon index. As with regions A and B, the abundances of Mg ($4 \sigma$ depletion) and Si ($\sim 3 \sigma$ depletion) are depleted relative to typical LMC values.

\begin{figure}
  % \plotone{f14.eps}
\plotone{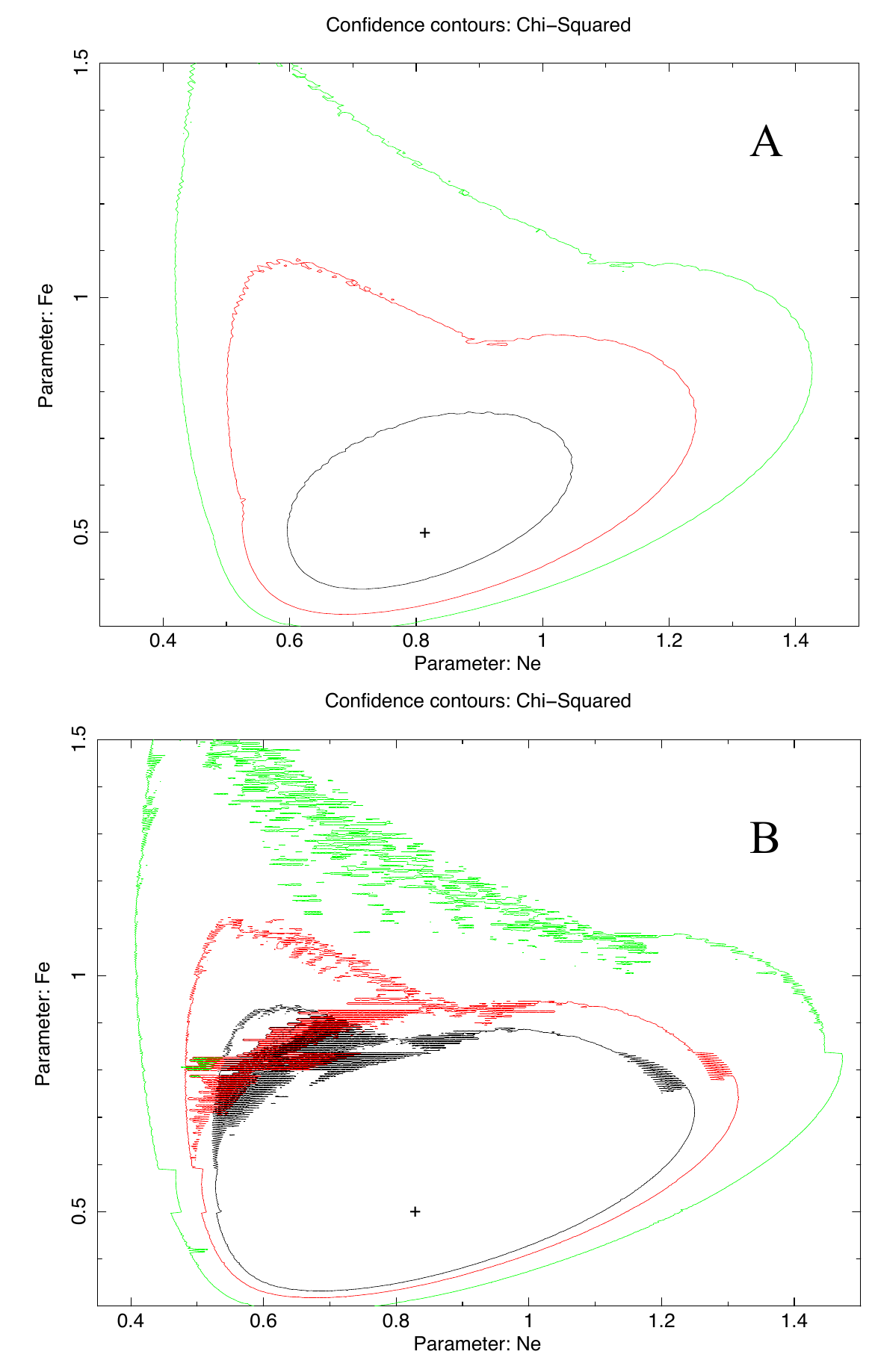}
\caption{Confidence-contour plot showing the variation in $\chi^2$ as the Ne and Fe abundances are varied in Region D for \textit{vnei} (A) and \textit{vpshock} (B) models. Contours are drawn around regions of 1$\sigma$, 2$\sigma$, and 3$\sigma$ change in the fit statistic as the two parameters are varied. The best-fit value of Ne and Fe abundances for each model is marked by a cross.}
\label{fig:D_ne_fe_conf}
\end{figure}

Region D covers a bright linear feature at the western end of the SNR shell, north of the synchrotron-emitting ``SW'' region analyzed by \citet{ParkHughes2010}. The most notable feature of this region is the overabundance of Ne ($A_{Ne}=0.8 \pm 0.1$, $\sim 4 \sigma$ above the typical LMC value). Ne and Fe lines, particularly in low-resolution spectra such as this, can be difficult to distinguish in thermal plasmas, so we generate similar confidence contours (Figure \ref{fig:D_ne_fe_conf}) for varying values of the Ne and Fe abundances to confirm that the Ne overabundance is real and not a result of degeneracy in the model parameter space. The contours show that for both \textit{vpshock} and \textit{vnei} fits the best-fit region for the neon abundance lies above the typical LMC value of $A_{Ne}=0.42$ and is well bounded on the lower end, while the iron abundance is consistent with the typical LMC value of $A_{Fe}= 0.5$.

Regions E and F cover separate portions of a bright knot in the southwest, near the ``SW'' region of \citeauthor{ParkHughes2010}.  Despite their physically distinct appearance, the two regions are both fit by an NEI model similar to that seen elsewhere in the shell, with temperatures of $kT \simeq 0.65$ keV and depleted abundances of Si ($\sim 6 \sigma$ below typical LMC value) and Mg ($\sim 3 \sigma$ below the typical LMC value). As with region C, the spectra of these regions do not show the higher-energy tail of Regions A, B, and D, and a second component is not required by the best-fit model in either region. 

\begin{figure}
  % \plotone{f15.eps}
  \plotone{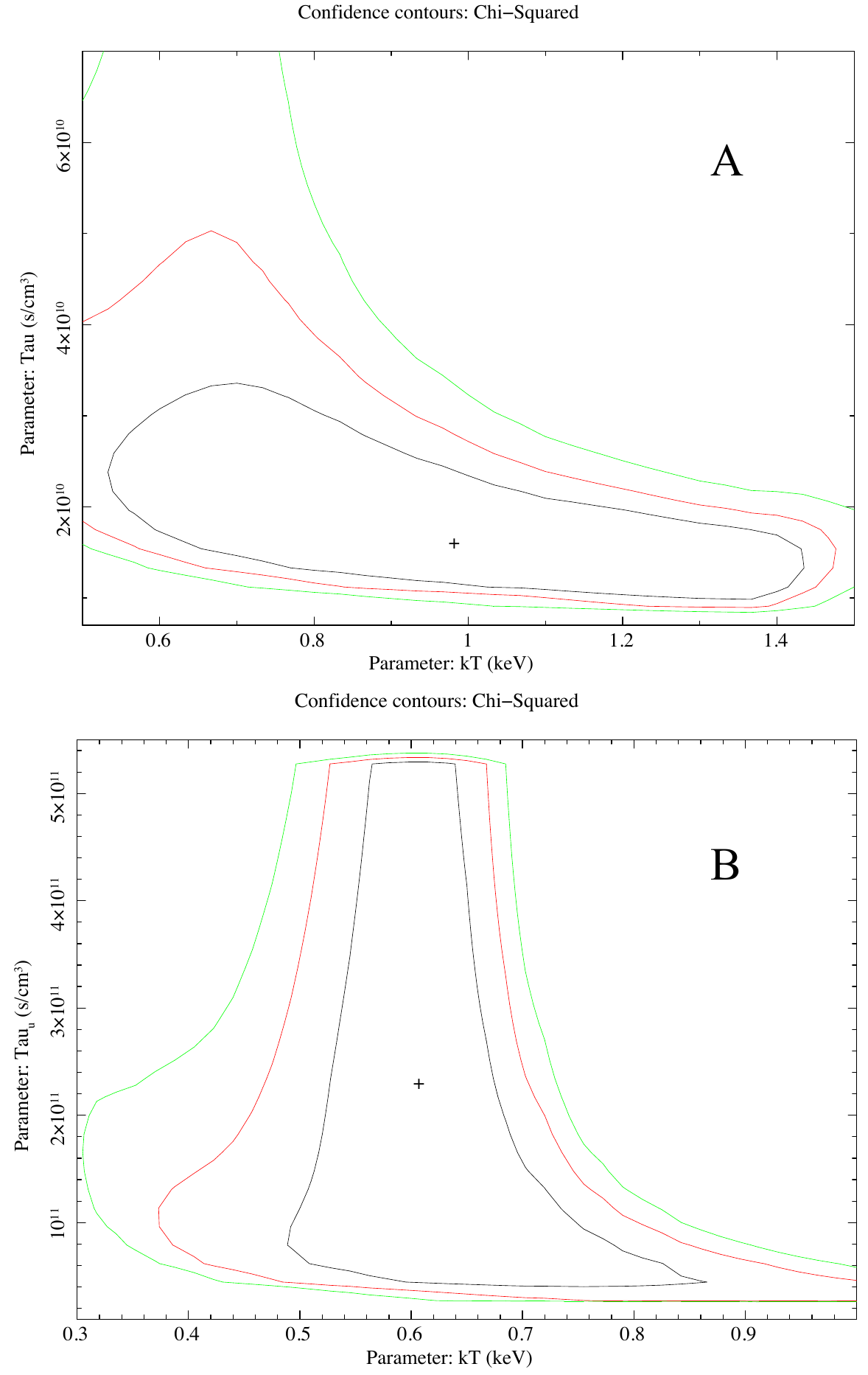}
  \caption{Confidence-contour plot showing the variation in $\chi^2$ as temperature and ionization timescale are varied in Region A for \textit{vnei} (A) and \textit{vpshock} (B) models. Contours are drawn around regions of 1$\sigma$, 2 $\sigma$, and 3$\sigma$ change in the fit statistic as the two parameters are varied. The best-fit value of $kT$ and $\tau$ for each model is marked by a cross.}
  \label{fig:A_kt_tau_conf}
\end{figure}

It is interesting to note that fitting the shell regions with \textit{vnei} and \textit{vpshock} models results in slightly different fits, as shown in Table \ref{tbl:nei_fits}. Although these two models both simulate NEI conditions, they do so differently; the \textit{vnei} model implements an NEI plasma with a single, uniform ionization parameter, while the \textit{vpshock} model implements an NEI plasma with a range of ionization parameters, with an upper bound of the fitted value of $\tau$. In general, \textit{vnei} and \textit{vpshock} fits result in almost identical values for the temperature and abundance, which is unsurprising. The exception is region A, where the two models disagree on the best fit temperature, and the error bars on the \textit{vpshock} model are unexpectedly small. We investigate these models by generating confidence-contour plots for varying values of temperature and ionization timescale. These plots, shown in Figure \ref{fig:A_kt_tau_conf}, reveal that the best-fit $kT$ regions for the two models overlap around $kT \sim 0.6-0.8$ keV. The differing best-fit temperatures for the two models are likely a result of the very poorly constrained ionization timescale in the \textit{vpshock} model, versus the \textit{vnei} model's relatively well-constrained ionization timescale.

The best fit values for ionization timescale differ greatly between \textit{vnei} and \textit{vpshock} models, and for most of the regions one of the two models does not have a well-constrained ionization timescale. This suggests that in regions A, B, E, and F, where the \textit{vnei} provides a better bound on the ionization timescale, the thermal X-ray emitting material is dominated by one specific density and hence one specific ionization timescale, so a single ionization parameter is a good approximation. For regions C and D, where a \textit{vpshock} model provides a better bound on the ionization timescale, we can infer that a wider distribution of densities and thus ionization timescales exists within the thermal plasma. 

We attempted to fit the higher-energy tail extending to $\sim 5$ keV in regions A, B, and D with a second thermal NEI component. For region A, the second thermal component requires a very high value for the ionization timescale ($\tau \sim 5 \times 10^{13}$) which is physically unrealistic for a nonequilibrium plasma \citep{SmithHughes2010}, but a second thermal component consisting of an equilibrium plasma results in an inferior fit statistic ($\chi^2_r=100/92$). For regions B and D, a two-temperature thermal model results in a fit statistic which is inferior to that of a thermal+power law or \textit{srcut} fit ($\chi^2_r=85/52$ for region B and $\chi^2_r=69/65$ for region D). In addition, for all three regions, the \textit{srcut} model results in a radio spectral index and 1 GHz flux consistent with our radio results reported in Section \ref{sec:radio_spec_index}. We therefore consider the higher energy emission in all three regions to be nonthermal in character.

The higher energy tails can be well-fit by a power law added to the NEI model, as is shown in Table \ref{tbl:nonthermal_fits}. The power-law indices for all three are consistent with values expected in synchrotron radiation from an SNR shock front \citep{Reynolds2008}. However, a pure power-law is a somewhat oversimplified model for synchrotron radiation. We thus attempt an additional fit by replacing the power law component in the fits with a cutoff synchrotron spectrum using the \textit{xssrcut} model. This model has three parameters: the radio spectral index $\alpha$, the radio flux density at 1 GHz, and the rolloff frequency. Although we have radio measurements of the flux density and the spectral index integrated over the entire shell, our observations, as detailed in Section \ref{sec:radio_spec_index}, do not allow us to measure these on spatial scales as small as our extraction regions. We thus allow all three parameters to vary freely when fitting. As shown in Table \ref{tbl:nonthermal_fits}, the radio spectral indices for all three regions are consistent with the integrated value of $\alpha =0.55$ found from the radio. The \textit{srcut} fit demonstrates that the data are consistent with synchrotron radiation from relativistic particles at the SNR shock front.

We can also calculate the maximum particle energy from the rolloff frequency as \citep{Reynolds2008}
 \begin{equation}
 E_{max} = 39 \left(\frac{h\nu_\text{rolloff}}{1\text{ keV}}\right)^{1/2} \left(\frac{B}{10 \text{ $\mu$G}}\right)^{1/2} \text{ TeV}
 \end{equation}
 As shown in Table \ref{tbl:nonthermal_fits}, assuming that the shell edge has a magnetic field on the order of $B = 10$ $\mu$G, the maximum particle acceleration energies are in the range $10-30$ TeV for these regions, although the errors are again large. The values for rolloff frequency and maximum particle energy are comparable to those seen in synchrotron-emitting shocks in other similarly-aged SNRs, such as RCW 86 ($\nu_\text{rolloff} = 0.8-1 \times 10^{16}$ Hz, which implies a maximum energy $E_{max}=20-25$ TeV; \citealt{RhoDyer2002}) and SN 1006 ($E_{max} = 20-80$ TeV; \citealt{RothenflugBallet2004}).

\subsection{Interior Regions}
\begin{figure}
  % \plotone{f16.eps}
  \plotone{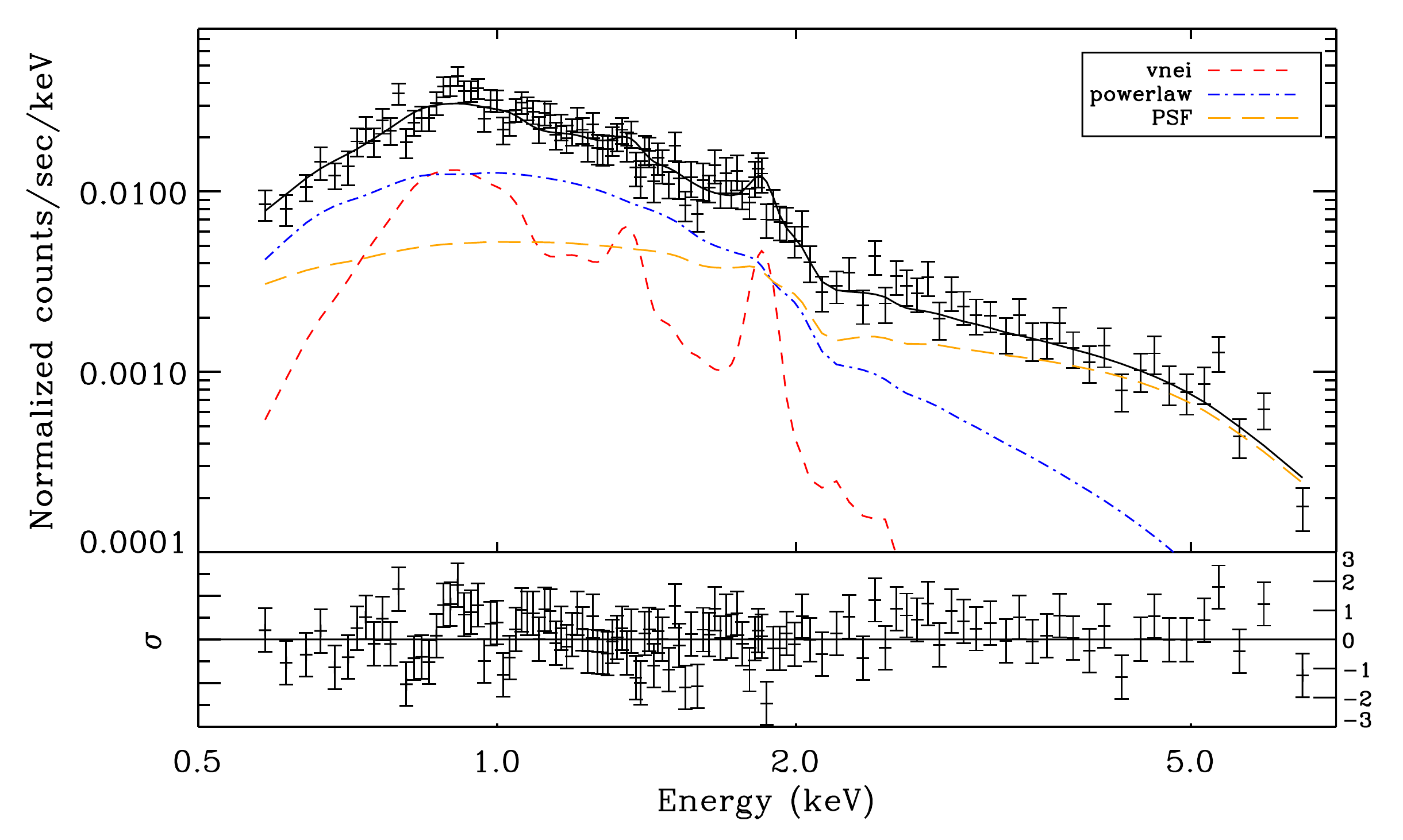}
\caption{Spectrum with best-fit model for the wedge centered at phase angle 0 (due west), shown as a representative spectrum of the wedge-shaped extraction regions in the halo. The contamination due to the pulsar's PSF is shown in orange.}
\label{fig:wedge_spec}
\end{figure}

Fit parameters for the wedge regions are shown in Table \ref{tbl:interior_fits}. These regions are all best fit by a \textit{vnei}+\textit{powerlaw} model. Although the ionization timescale for most of the fits is large and/or not well constrained, a \textit{vnei} model results in a superior fit statistic to an equilibrium (\textit{vapec}) model for all regions. We note that although the power-law continuum is the dominant feature in the spectrum, the thermal component has a minor but crucial contribution to the fit, and pure power law models do not adequately fit the spectra. A representative spectrum for the wedge regions can be seen in Figure \ref{fig:wedge_spec}. Analysis of the spectra from the halo region is affected by severe contamination from the PSF of the pulsar and PWN, which can be seen in the sample spectrum in Figure \ref{fig:wedge_spec}. The best fit values for each wedge have some slight variation across different azimuthal angles, but given the large errors on these fits, are mostly consistent with each other.

\section{Discussion}\label{sec:discussion}
Having dealt with the analysis of radio and X-ray data separately in preceding sections, we now combine results from all surveyed data to form a physical picture of 0540. In Section \ref{sec:shell_discussion}, we use calculated shock parameters from the X-ray spectra together with magnetic field and Faraday rotation measure information from the radio observations. Together with morphological considerations from all surveyed wavebands, we use this information to to describe and elucidate the physical characteristics of the shell and its interaction with the surrounding ISM.
 In Section \ref{sec:halo_discussion}, we discuss the X-ray halo surrounding the PWN and, based on X-ray spectroscopy and morphological considerations from optical and X-ray, show that it is likely caused by foreground shell emission seen in projection.

\subsection{Shell} \label{sec:shell_discussion}

\subsubsection{Density and Shock Velocity}\label{sec:shell_density}
The shell morphology, as discussed in Section \ref{sec:morphology}, is indicative of an interaction between the shell and surrounding denser clouds. We can elucidate this interaction by examining densities and shock velocities calculated from the X-ray spectra.
All surveyed regions in the shell show spectra that are best fit by NEI conditions. The regions at the outer extent of the X-ray emission (A, B, and D) have a power-law component added to this soft thermal component. The softer thermal emission is likely caused by the interaction between the expanding supernova blast wave and the surrounding ISM. We can calculate the approximate blast wave speeds in these regions by solving the Rankine-Hugoniot relations in the case of a strong shock, with $\gamma = 5/3$, which yields the equation $v=\sqrt{16kT/3\mu m_p}$, with $\mu = 1.2$ for a fully ionized plasma. This relation assumes that the electron and ion tempertures have equilibrated, which is not necessarily the case (e.g. \citealt{GhavamianLaming2007}). However, in a fully ionized, shock-heated plasma, the electron and ion temperatures can equilibrate very quickly (within a few hundred years; see Eq. 36.37 in \citealt{Draine2011}) via Coulomb interactions. We may thus use this equation as a conservative lower limit.

The resulting velocities, which can be seen in Table \ref{tbl:shock_speeds}, are on the order of a few hundred km s$^{-1}$. This is extremely slow for a SNR as young as 0540. We may conclude from this that the outward-moving blast wave has interacted with denser surrounding material fairly recently, and that it has quickly decelerated. We calculate the density based on the best-fit ionization timescale $\tau$ for each region; the density is thus $n_e = \tau / t$, where $t$ is the time since the region has been shocked. This time, $t$, is estimated by assuming the initial blast wave speed is $\sim 10,000$ km s$^{-1}$ \citep{Chevalier1977} and that the initial SNR blast wave is spherically symmetric. We assume that the blast wave expands at this constant speed until it reaches the position of the shell region in question. We then calculate $t$ for each region by subtracting the time it takes the blast wave to reach that region from the accepted age of 1100 years. The densities calculated in this way for each region are listed in Table \ref{tbl:shock_speeds}.

We can check these results of this calculation by calculating the density in a different way, based on the volume of the emitting region and the emission measure ($\int n_e n_h dV$) of the thermal plasma in that region. These quantities can be calculated from the thermal normalization parameter. If we assume that the densities of electrons and protons are roughly constant through the volume in each region and that $n_e = 1.2 n_h$, we can then estimate the density as
\begin{equation}
  \label{eq:density_from_norm}
  n_e = (\frac{4 \pi N_a}{1.2 V f}D^2_{cm})^{1/2}
\end{equation}
where $N_a$ is the thermal normalization parameter, $D_{cm}$ is the source distance in cm, and $V$ is the volume of the emitting region in $cm^3$. The volume filling factor $f$ is a number between 0 and 1 to account for the percentage of the volume in each region actually taken up by emitting plasma. We assume that the source distance is 50 kpc and that the volume of each region can be approximated by its 2-dimensional area on the plane of the sky multiplied by the longest axis of the region.

The results estimated from the normalization parameter, shown in Table \ref{tbl:shock_speeds}, are roughly consistent with the values estimated from the ionization timescale, except for the value seen in region D, where the density calculated from the NEI norm is nearly an entire order of magnitude lower. It is possible that the initial shock speed assumed in the calculation in Table \ref{tbl:shock_speeds} is too high, and that the shock front thus reached this region more recently than assumed in this calculation. It is also possible that the volume filling factor is unusually low in this region. Finally, it is possible that the estimated volume for this region is incorrect. The shape of the western shell in 0540 is complex and very non-spherical, making an accurate estimate of the emitting volume of each region difficult. 

We can see that in higher density regions, lower velocity shocks predominate and conversely, in lower density regions, higher velocity shocks predominate. For instance, Region D, with the highest density of any surveyed shell region, shows the lowest temperature X-ray plasma and the slowest shock. This is precisely the situation we would expect to see if the shock was interacting with a dense cloud in the surrounding ISM.

\subsubsection{Multi-wavelength Emission}
\begin{figure}
  % \plotone{f17.eps}
  \plotone{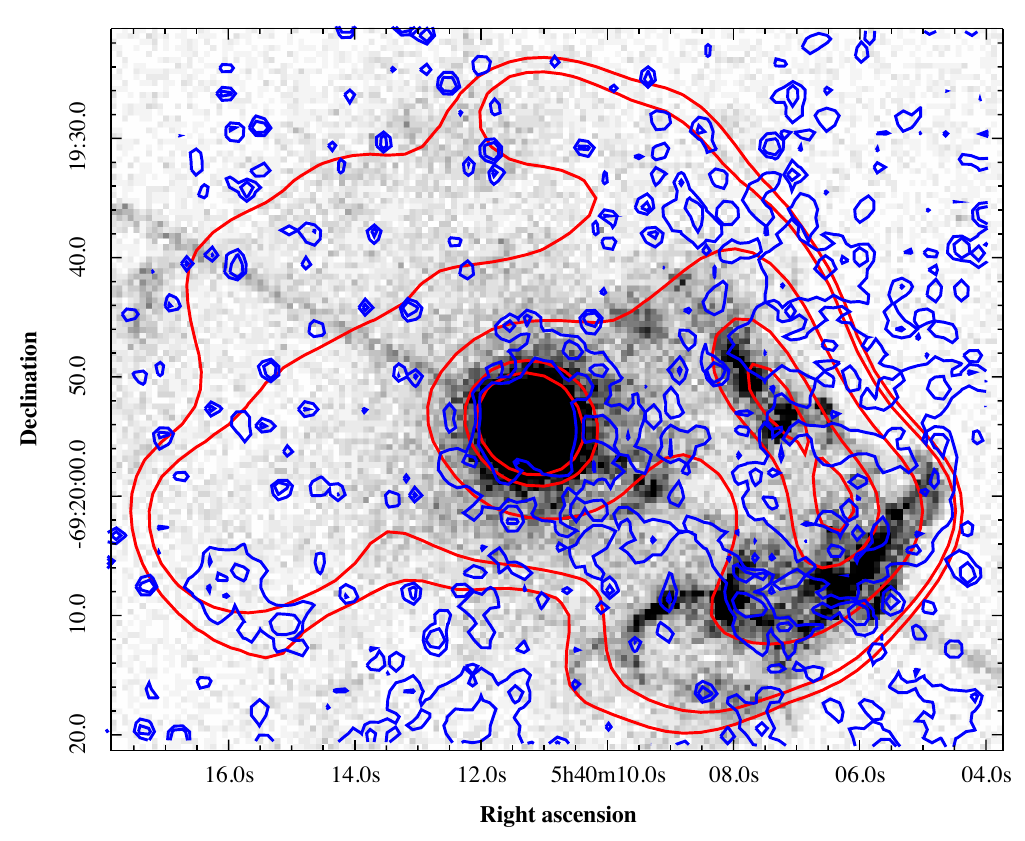}
\caption{\textit{Chandra} X-ray data in reverse gray-scale, unbinned and scaled linearly from 0-75 counts per bin. Contours for [\ion{O}{3}] are drawn in blue at levels of 0.06 and 0.09 ADU s$^{-1}$ and contours for 1513 MHz radio emission are drawn in red at levels of 7.5, 10, 25, 60, and 80 mJy beam$^{-1}$.}
\label{fig:CXO_gray_contours}
\end{figure}

The highly inhomogeneous density of the ISM surrounding 0540 also results in a variety of types of emission around the edge of the shell. 
As shown in Figure \ref{fig:CXO_gray_contours}, Regions B and D are coincident with an area of diffuse [\ion{O}{3}] emission \citep{MathewsonFord1983} and with knots of radio emission. The multi-wavelength emission in the western hemisphere is indicative of a blast wave expanding into a complicated medium, with the radio knots and synchrotron emission seen in the X-ray spectra representing areas of lower density and the softer thermal X-rays and optical emission representing interactions with density enhancements in a clumpy medium. In addition, longer wavelength IR emission in the southwest appears outside of the X-ray emission (see Figure \ref{fig:Spitzer_tricolor}), suggesting that a particularly dense cloud has been shock heated with only larger dust grains remaining. The shorter grains have been destroyed by the shock front, or sputtered away in the hot, shocked material, causing the lack of shorter-wavelength diffuse IR emission discussed in Section \ref{sec:IR_obs}. This cloud may be leftover material from the progenitor star's formation, sitting at the edge of a cavity swept out by the progenitor's wind.

The western area of the shell also shows strong variations in rotation measure in Figure \ref{fig:FaradayRM}, such as the knots of reversed rotation measure around $\alpha=5$h 40m 7s, $\delta=-$69\textdegree\ 19\arcmin\ 58\arcsec, which suggest that the electron density and magnetic field direction in this region varies on relatively small scales, as we would expect in a highly clumpy medium. On a larger scale, the magnetic field vectors along the entire western edge of the shell, shown in Figure \ref{fig:B_field}, do not appear to be oriented radially, as is typical of young SNRs \citep{Milne1987}. The magnetic field vectors instead appear to be tangential along the edge of the radio emission, a configuration usually seen in older SNRs and attributed to compression in shocks that have decelerated into the radiative regime \citep{ReynoldsGaensler2012}.

The presence of a hard nonthermal component in regions A, B and D, which is evident in Figure \ref{fig:shell_spectra}, is likely due to the presence of synchrotron radiation, which was also seen at the extreme east and west edges of the shell by \citet{ParkHughes2010}. The X-ray photon indices in these regions of $\Gamma \sim 3$ are within the range of other young SNR with synchrotron emission at their shock fronts \citep{Reynolds2008}, and the results of an \textit{srcut} fit to the nonthermal component further support the idea of synchrotron emission in these regions. Unfortunately, due to observational constraints detailed in Section \ref{sec:radio_spec_index}, calculating the radio spectral index for these regions is not feasible. The superimposition of synchrotron emission and thermal emission may be due to line of sight confusion along the clumpy, highly inhomogeneous medium. In lower density areas the shock is collisionless and able to sweep up the ambient magnetic field and accelerate particles to relativistic velocities, thus producing the observed synchrotron emission. Shocks that generate synchrotron emission in this manner are usually very fast ($>3000$ km s$^{-1}$ or more; \citealt{Bell1978a}), yet we see the fast, collisionless shocks required for synchrotron emission superimposed on the much slower shocks producing thermal emission. This requires that the slower thermal shocks have decelerated very recently - for instance, for a $\sim 3000$ km s$^{-1}$ synchrotron producing shock and the 400 km s$^{-1}$ thermal shock to both appear in Region D, which is 1 pc thick, the slower shock must have decelerated to its current speed within the last $\sim 350$ years. 

\subsubsection{Comparison with the Crab Nebula}
This interaction of a shell with a dense surrounding medium is less apparent in 0540's ``twin", the Crab. The lack of a visible shell in the Crab has been attributed (\citealt{Hester2008} and references therein) to the explosion occurring within a low-density cavity blown out by the wind of the progenitor, which likely had a mass of $8-13 M_{\odot}$. 0540 shows signs of having exploded within a similar cavity, but its blast wave has recently encountered the clumpy, dense material at the western edge, while remaining relatively unimpeded in the east. The superbubble N158 is located less than 2\arcmin\ to the south of 0540 \citep{DunnePoints2001}. This bubble is expanding outward \citep{SasakiBreitschwerdt2011} and driving a shock into the surrounding ISM, which strongly influences the dynamics of the area around it. The cavity blown out by the progenitor of 0540 might be subject to crushing from the surrounding bubbles as they expand into it. A type IIP explosion, which has been suggested for 0540 \citep{Chevalier2005}, would imply a red supergiant (RSG) progenitor, which would have a weak wind. The combination of a weak stellar wind and a complex environment is contrary to the Crab's low-density cavity, which could help explain the complex shell morphology in 0540.

\subsubsection{Cloud/Shock Interaction}
We see a distinctly higher temperature in Region C than in the other surveyed shell regions. The combination of higher temperature, substantially lower density, and shorter ionization timescale than Region D, immediately exterior to it, suggests that this region has been recently shocked and reheated by a reflected shock. The extremely high density in Region D ($n_e$ as high as $\sim 30$ cm$^{-3}$ or more), immediately exterior to Region C ($n_e \sim 3$ cm$^{-3}$) should have caused the blast wave to decelerate quickly upon reaching the denser material. The abrupt deceleration of the blast wave in Region D may have caused a prominent reflected shock to begin propagating back towards the center of the remnant, reheating interior regions such as Region C.

The depleted Mg and Si abundance seen in nearly all shell regions surveyed may be related to silicate dust condensation behind the shock, which would emit infrared radiation. However, the partial shell-like structure in the IR, visible in Figure \ref{fig:Spitzer_tricolor}, is immediately outside the X-ray contours, suggesting that it is caused instead by emission from dust leftover from the progenitor's formation, shocked by the expanding SNR blast wave. This structure only appears in the two longer-wavelength IRAC bands (5.8 and 8 $\mu$m), suggesting an absence of small dust grains. The larger dust grains that emit in these longer-wavelength bands have a longer sputtering lifetime and have not yet evaporated in the hot, shocked material. If the depleted Mg and Si abundances in the shell are not due to dust formation, the region surrounding 0540 may simply be locally depleted in these elements. A photoionization study of a bright filament within 30 Doradus \citep{TsamisPequignot2005} found evidence for regions of depleted elemental abundances on small scales. These authors examined optical and IR spectra of a bright filament and found severely depleted Mg and Si abundances (Mg=0.13 solar, Si=0.09 solar). It is therefore quite possible that 0540 exploded in a region of similarly depleted Mg and Si. 

Although the interaction between the SNR blast wave and the shell has progressed to the point where much of the X-ray emission is dominated by shocked ISM swept up by the blast wave, rather than ejecta from the progenitor, ejecta clumps may still exist. The overabundance of Ne in Region D may indicate that the emission in this region includes a clump of shocked ejecta from the progenitor star. From the densities calculated in Section \ref{sec:shell_density} and shown in Table \ref{tbl:shock_speeds}, we can calculate the total emitting mass of Ne. The estimated density for Region D is  $3f^{-1/2}-30$ cm$^{-3}$, the total emitting mass of this region is $ 0.6f^{-1/2}-6 M_{\odot}$. The mass fraction of Ne in this region is, from the best-fit abundance, $X_{Ne}=9.9 \times 10^{-5}$, which implies an emitting mass of Ne of $\sim 6 \times 10^{-4}- 6f^{-1/2} \times 10^{-5} M_{\odot}$. For models of $15-25$ $M_{\odot}$ SNe with sub-solar metallicity \citep{WoosleyWeaver1995}, the expected yield is about $5 \times 10^{-2} M_{\odot}$ of Ne, so we may be seeing a ``bullet" of ejecta comprising about $0.1f^{-1/2}- 1$\% of the Ne produced in the explosion, similar to those seen in Cassiopeia A \citep{WillingaleBleeker2002} or Vela \citep{WangChevalier2002}. We may add this to the tentative detection of Fe-rich material in the south by \citet{ParkHughes2010} and to the high pulsar kick velocity (1190 km s$^{-1}$) measured by \citet{SerafimovichShibanov2004}, which appears to be pointing to the south, more or less along the pulsar's spin axis. This evidence -- the highly asymmetric distribution of the ejecta and the large pulsar kick velocity -- points to the collapse and explosion of the progenitor star being strongly asymmetric, which has likely contributed (along with the anisotropy of the surrounding ISM) to the asymmetry of the shell.

\subsubsection{Shell Summary}
The overall picture that we see in the shell can be summarized as follows. The western region of 0540 is bright in all wavelengths. The emission is caused by the blast wave having interacted recently with dense, clumpy material. The western side of 0540's shell contains ambient dense clouds that have most likely remained in the environment since the progenitor star's formation. There is no evidence of such features in the eastern half of the remnant. It is possible that local bubbles such as N158 are impeding the dissipation of the material in the west or, alternatively, assisting with rarefication in the east. The shocked ISM displays depleted Mg and Si abundances, possibly due to dust depletion or locally depleted abundances of these elements. There is evidence that a reverse shock has begun propagating back towards the center of the SNR, producing higher temperature nonequilibrium conditions in lower-density media.  The soft thermal X-rays from the SNR shell are superimposed on synchrotron emission from more tenuous regions due to projection along the line of sight. As shown in Figure \ref{fig:CXO_gray_contours}, this superimposition extends to [\ion{O}{3}] and radio emission as well, which appear together with thermal and nonthermal X-rays in the complex western edge of the shell. The radio spectral indices support the presence of synchrotron radiation. The large variation seen in rotation measure along the western edge of the shell in Figure \ref{fig:FaradayRM} supports the idea of mixing that has strongly affected the electron density distribution.

\subsection{Interior Halo}\label{sec:halo_discussion}
The X-ray observations of the halo region show a mixture of thermal and nonthermal emission. Several explanations have been advanced for the cause of this diffuse halo emission. It is possible that Rayleigh-Taylor (RT) instability between the expanding plerion and the surrounding ejecta may lead to mixing between the synchrotron emitting PWN material with thermal plasma. It is also possible that the diffuse halo emission could be caused by photons from the pulsar and PWN being scattered by foreground dust. A dust halo of this type is seen in the galactic SNRs G$21.5-0.9$ \citep{Bocchinovan-der-Swaluw2005,MathesonSafi-Harb2010} and Kesteven 75 \citep{HelfandCollins2003}. \citet{MorseSmith2006} have suggested, based on optical line profiles, that the diffuse [\ion{O}{3}] emission in the halo is caused by the forward shock from the expanding PWN overtaking slow-moving ejecta in the interior of the SNR. Alternatively, \citet{WilliamsBorkowski2008} have suggested, based on IR spectra, that the diffuse halo emission is caused by UV photons from the PWN photoionizing material immediately surrounding the nebula. Finally, \citet{PetreHwang2007} have noted that the halo emission may be simply foreground material from the SNR shell seen in projection.

Rayleigh-Taylor (RT) instabilities, arising from the expansion of the pulsar wind into the denser surrounding environment, can cause mixing between thermal and synchrotron emitting material, such as that seen in the older LMC PWN 0453-68.5 \citep{McEntafferBrantseg2012}. In that remnant, RT instabilities are caused by repeated compression and expansion of the PWN due to interactions with the SNR reverse shock. However, RT instability can also arise from the expansion of the tenuous synchrotron-emitting PWN material into the denser surrounding ejecta early in the life of a PWN \citep{Jun1998}. Noting that thermal material and power-law emitting material are seen together from radii of 5\arcsec\ to at least 9\arcsec\ ($1.2-2.2$ pc from the pulsar), we can calculate the requirements for RT instability to occur on this scale. \citet{JunNorman1995} give the most rapidly growing and therefore dominant length scale $\lambda_c$ for RT instability, given a magnetic field at a set strength $B$, as
\begin{equation}
\lambda_c = \frac{2B^2 \cos^2 \theta}{g(\rho_2 - \rho_1)}
\end{equation}
where $g$ is the local acceleration (effective gravity), $\theta$ is the angle between the field and the interface surface, and $\rho_2$ and $\rho_1$ are the densities of the lighter (PWN) and heavier (SN ejecta) media. The field should be parallel to the interaction surface everywhere ($\cos \theta \sim 1$), as suggested by polarization studies of the nebula (\citealt{ChananHelfand1990,LundqvistLundqvist2011}; Figure \ref{fig:B_field}). We can calculate the required density differential \citep{HesterStone1996} as
\begin{equation}
\delta \rho = \frac{B^2}{\lambda g}
\end{equation}
This is a minimum density jump for RT instability to occur at the interface, below which the magnetic field stabilizes the interface and no mixing of any sort occurs. For a density jump $\sim \delta \rho$, the magnetic field shapes the material at the interface into long fingers \citep{Jun1998}; if the density jump is $\gg \delta \rho$, the effect of the field is unimportant and we recover the purely hydrodynamic case. The effective gravity of the similar synchrotron nebula in the Crab is $\simeq 3.53 \times 10^{-3}$ cm s$^{-2}$ \citep{HesterStone1996}. 

Using the PWN magnetic field value of 250 $\mu$G calculated in Section \ref{sec:broadband_spec}, we get an estimated critical density jump of $3.8 \times 10^{-25}$ g cm$^{-3}$, or about 0.2 cm$^{-3}$ for inner ejecta with an average atomic mass $\mu = 1.2 m_H$.  Estimating the density of the thermal plasma in the wedge regions leads to a number density in this region of $\sim 2/f$ cm$^{-3}$. As this is at least ten times the critical density, and, depending on the value of the filling factor, perhaps well over an order of magnitude higher, it is at least theoretically possible that RT instability could be the cause of the power-law tail seen in the halo spectra. 

To check the dust-scattering model, radial brightness profiles for the hard ($1.2-8.0$ keV) emission were also extracted from eastern and western halves of the interior area between radii of 10 and 60 pixels ($5-30$\arcsec) using the \textit{Sherpa} tool \texttt{prof\_data}.  To remove the contribution to this halo from the \textit{Chandra} PSF, we used our MARX simulations to extract brightness profiles for the PSF from the pulsar and PWN and subtracted these simulated brightness profiles. The PSF-subtracted profiles were fit to a simple dust-scattering halo model \citep{Hayakawa1970} of the form $I \propto [j_1(x)/x]^2$, where $j_1(x)$ is the spherical Bessel function of the first kind. The  profile for the western half becomes non-monotonic about 35 pixels (18\arcsec) from the pulsar, so we cut off the fit at that point. We can confidently reject foreground dust scattering as a primary cause for the halo based on several considerations. In the $1.2-8.0$ keV band, where dust scattering would be most noticeable, the brightness profile is not well fit at all ($\chi^2_r \sim 9$) by the dust scattering model. We may also note, qualitatively, that dust scattering halos of this type are mainly observed around objects in the galactic plane with very high column densities. For instance, G$21.5-0.9$ has a column density of $n_H = 2.2 \times 10^{22}$ cm$^{-2}$ \citep{MathesonSafi-Harb2010}, four times that of 0540. In addition, the gas-to-dust ratio in the LMC is more than double that of the Milky Way \citep{MeixnerGalliano2010}, further reducing the scattering contribution from LMC dust. Given the low column density through the Milky Way to 0540, we do not expect any significant amount of scattering from local galactic dust. 

The shock \citep{MorseSmith2006} and photoionization \citep{WilliamsBorkowski2008} scenarios for the halo emission each rely on the presence of ejecta surrounding the PWN. In [\ion{O}{3}] images, the PWN is surrounded by a prominent ring-like structure. A similar [\ion{O}{3}] ring structure exists around the Crab Nebula, which has been interpreted \citep{SankritHester1997} as cooling immediately behind a shock being driven into the inner ejecta by the expanding PWN. \citet{MorseSmith2006} have suggested that the halo may be the result of a fast shock from the expanding PWN overtaking and heating inner, slow-moving ejecta. \citet{KirshnerMorse1989} examined the ratio of the [\ion{S}{2}] $\lambda \lambda 4069,4076$ doublet to the [\ion{S}{2}] $\lambda \lambda 6717,6731$ doublet and found that the ratio was consistent with a temperature of about 10,000 K, characteristic of S$^+$ in shock heated gas \citep{FesenBlair1982}.
In the photoionization scenario, supported by \textit{Spitzer} IR spectra \citep{WilliamsBorkowski2008}, a very slow shock of about 20 km s$^{-1}$ is currently propagating outward inside the 8\arcsec\ halo, while a faster shock of $\sim 250$ km s$^{-1}$ is propagating outward outside the halo. The authors imply the low shock velocity from the emissivity of lines such as [\ion{O}{1}], [\ion{O}{4}], [\ion{Fe}{2}], and [\ion{S}{2}]. These authors argue that the slow shock is best explained by confinement within an iron-nickel bubble \citep{Woosley1988,LiMcCray1993}. In this phenomenon, which has also been observed in SN1987A \citep{Basko1994,Wang2005}, radioactive energy deposition by the decay of $^{56}$Co and $^{56}$Ni causes an inner bubble of heavy-element ejecta to expand rapidly and sweep up a clumpy shell of inner ejecta. UV photons emitted from the PWN photoionize this surrounding ejecta, which is slow-moving and highly enriched ($A_{metal} \simeq 2$) in heavy elements, to produce the observed [\ion{O}{3}] emission. The X-ray spectra for the halo emission are characterized by a plasma temperature of $kT \sim 0.7$ keV, which is consistent with a shock moving at $\sim 550$ km s$^{-1}$, but if this shock were moving through heavy-element ejecta, we would expect the X-ray spectra to show enhanced elemental abundances from this shock-heated ejecta, which they do not.

Neither the shock, photoionization, or RT scenarios present an entirely satisfying explanation of the X-ray emission. The lack of X-ray ejecta emission does not present a problem in the RT scenario, since the analogous RT filaments in the Crab do not emit in the X-ray \citep{WeisskopfHester2000}. However, we do not see visible RT filaments at any wavelength in the halo region. Filamentary structures that may be due to RT instability are present within the PWN, as shown by HST WFPC observations in [\ion{O}{3}] and [\ion{S}{2}] \citep{MorseSmith2006}, but these filaments do not extend beyond the PWN. These difficulties with explaining the X-ray emission from the halo leave us with the possibility suggested by \citet{PetreHwang2007} that the halo emission may merely be a chance alignment of an area of brighter foreground emission from the shell. Despite the strong contamination, the thermal component best-fit X-ray model is almost identical in temperature and ionization timescale to those seen in the shell. This supports the explanation of line of sight confusion between the shell and the plerion. 

\section{Conclusions}
We have undertaken a study of emission from the shell and interior of the young composite supernova remnant SNR 0540-69.3 involving detailed analysis of archival X-ray and radio data, together with supporting evidence from optical and IR data. We briefly summarize our results below:

\begin{enumerate}
\item The SNR shell is interacting with clumpy material that is highly inhomogenous in density, surrounding a low-density cavity blown out of the surrounding material by the supernova shock wave or the wind of the progenitor star. Thermal X-ray spectra from the shell indicate that the shock speeds at the western edge of the shell are far lower than expected for an SNR as young as 0540, suggesting the presence of very dense material in the west. The very hot thermal plasma in the north (Region A) and in the south and east \citep{ParkHughes2010} has not been affected by these clumps of dense material, suggesting that the surrounding ISM is very asymmetric. The correlation of [\ion{O}{3}] and X-ray emission and the prevalence of non-radial magnetic fields in portions of the shell, particularly at the western edge, also suggests the presence of very dense material. The presence of X-ray and radio synchrotron emission in portions of the shell also suggests that there are regions of far lower density. 
\item The asymmetry of the shell is caused by multiple factors. First, the presence of a possible Ne-rich ejecta knot, as well as the possible Fe-rich ejecta knot detected by \citet{ParkHughes2010} and the high pulsar kick velocity detected by \citet{SerafimovichShibanov2004}, indicates that there was a high degree of asymmetry in the initial collapse and explosion of the progenitor star. Second, the fact that lower shock speeds appear in regions of higher density, as shown in Table \ref{tbl:shock_speeds}, indicates that the outward expansion of the shell has been strongly affected by density inhomogeneities and clumps in the surrounding environment. Finally, the complexity of the surrounding environment, including various clouds and star-forming regions, may be creating, sustaining, or dissipating these density variations.
\item Although a diffuse halo is seen around the PWN in X-ray that appears similar to the extended emission seen in [\ion{O}{3}] images, the X-ray emission is probably caused by a chance alignment of foreground material in the shell and has no relation to the optical emission from the ejecta immediately surrounding the PWN. The explanations that have been advanced for the [\ion{O}{3}] and IR emission from this region do not provide a satisfactory explanation for the X-ray halo. In addition, the thermal X-ray emission in this region is very similar to other regions in the shell. We do not see evidence of any effect of the shell or its environment on the PWN or its immediate surroundings.
\end{enumerate}

The plerion in 0540 is usually considered to be a near-twin of the Crab. The two plerions are indeed quite similar, but the distinct shell and center-filled morphology in 0540 contrasts sharply with the invisible shell in the Crab. We see no evidence that any significant interaction or mixing between the ejecta and the PWN is taking place in 0540, as in the Crab, so a comparison between the two PWNe is reasonable at present. The differences in environment will, in time, cause the evolutionary paths of the two objects to diverge substantially. As the outgoing blast wave continues to interact with the ISM, the reflected shocks will become stronger and propagate back towards the center of the SNR. The complex, strongly asymmetric environment may cause the reverse shock to evolve in a similarly asymmetric manner, with potentially interesting effects on the future of the PWN in 0540. Although the Crab has a long history of observation along the entire electromagnetic spectrum, additional observations of 0540 could help to add to our understanding of the similarities and differences between the evolutionary paths of these two prototypical PWNe.

\section{Acknowledgements}
The authors gratefully acknowledge helpful commentary on the manuscript by Rob Petre, as well as many helpful suggestions and comments from the anonymous referee.
T. B. is supported by NASA grant NNX10AN16H. The results in this paper are based in large part on data obtained from the \textit{Chandra} Data Archive. This work is based in part on observations made with the Spitzer Space Telescope, which is operated by the Jet Propulsion Laboratory, California Institute of Technology under a contract with NASA.
This paper utilized observations made with ESO Telescopes at the La Silla Observatory under program ID 076.C-0888, processed and released by the ESO VOS/ADP group. This paper made use of the \textsc{karma} software package developed by the ATNF. The Australia Telescope Compact Array is part of the Australia Telescope, which is funded by the Commonwealth of Australia for operation as a National Facility managed by CSIRO.

%%%%%%%%%%%%%%%%%%%%%%%%%%%%%%%%%%%%%%%%%%%%%%%%%%%
%%%%%%%%%%%%%%%%%%%%% FIGURES %%%%%%%%%%%%%%%%%%%%%
%%%%%%%%%%%%%%%%%%%%%%%%%%%%%%%%%%%%%%%%%%%%%%%%%%%

%%%%%%%%%%%%%%%%%%%%%%%%%%%%%%%%%%%%%%%%%%%%%%%%%%%
%%%%%%%%%%%%%%%%%%%%% TABLES %%%%%%%%%%%%%%%%%%%%%%
%%%%%%%%%%%%%%%%%%%%%%%%%%%%%%%%%%%%%%%%%%%%%%%%%%%

\clearpage
\begin{deluxetable}{lccc}
\tablecaption{Summary of ATCA observations used}
\tablehead{
	\colhead{Date} &
	\colhead{Scan time} &
	\colhead{Array} &
	\colhead{Frequencies}\\
	& (minutes) & & (MHz)
	}
\startdata
2000-Jun-03 & 553 & 6B & 1344,1728\\
2000-Jun-02 & 544 & 6B & 1344,1728\\
1995-Oct-24 & 298 & 1.5D & 5824,8896\\
1995-Oct-23 & 1164.5 & 1.5D & 5824,8896\\
1992-Jun-26 & 587.5 & 6A & 4790,8640\\
1992-Jun-25 & 460 & 6A & 4790,8640\\
1992-Jul-15 & 354.5 & 6D & 4790,8640\\
1991-Sep-14 & 79.5 & 6C & 2290\\
1991-Sep-13 & 438.2 & 6C & 2290\\
\enddata
\label{tbl:ATCA_observations}
\end{deluxetable}

\clearpage
\begin{deluxetable}{ccccccc}
\tablecaption{Integrated radio flux density of SNR 0540-69.3}
\tablehead{
	\colhead{$\nu$} &
	\colhead{$\lambda$} &
	\colhead{RMS} &
	\colhead{Beam size} &
	\colhead{S$_{total}$} &
	\colhead{S$_{core}$} &
	\colhead{S$_{shell}$}\\
	(MHz) & (cm) & (mJy) & (\arcsec) & (mJy) & (mJy) & (mJy)
}
\startdata
843 & 36 & 0.8 & 46.4$\times$43.0 & 1147 & \nodata & \nodata\\
843 & 36 & 1.0 & 46.4$\times$43.0 & 1104 & \nodata & \nodata\\
1384 & 20 & 0.6 & 40.0$\times$40.0 & 835 & \nodata & 767\tablenotemark{a}\\
1513 & 20 & 2.0 & 9.7$\times$7.9 & \nodata & 68 & \nodata\\
2290 & 13 & 1.0 & 6.6$\times$4.9 & \nodata & 62 & \nodata\\
4790 & 6 & 0.4 & 3.4$\times$2.8 & \nodata & 54 & \nodata\\
4800 & 6 & 0.7 & 35.0$\times$35.0 & 411 & \nodata & 351\tablenotemark{a}\\
5824 & 6 & 0.6 & 5.2$\times$3.7 & \nodata & 60 & \nodata\\
8640 & 3 & 0.2 & 1.9$\times$1.5 & \nodata & 51 & \nodata\\
8640 & 3 & 0.7 & 22.0$\times$22.0 & 285 & \nodata & 234\tablenotemark{a}\\
\enddata
\tablenotetext{a}{These values were calculated and not directly measured. Details are given in Section \ref{sec:radio_spec_index}.}
\label{tbl:radio_flux}
\end{deluxetable}

\clearpage
\begin{deluxetable}{cccccccccccc}
\rotate
\tabletypesize{\scriptsize}
\tablecaption{Comparison of \textit{vnei} and \textit{vpshock} fits for Shell Regions}
\tablewidth{0pt}
\tablehead{
	\colhead{Region} &
	\colhead{Model} &
	\colhead{$\chi^2/\nu$} &
	\colhead{$n_H$} &
	\colhead{$kT$} &
	\colhead{Mg} &
	\colhead{Si} &
	\colhead{Ne} &
	\colhead{$n_e t$} &
	\colhead{norm} &
	\colhead{$\Gamma$} &
	\colhead{norm$_{PL}$}\\
	& & & ($10^{22}$ cm$^{-2}$) & (keV) & & & & ($10^{11}$ cm$^{-3}$ s) & $10^{-5} N_a$\tablenotemark{a} & & $10^{-5} N_b$\tablenotemark{b}
}
\startdata
A & \textit{vpshock}+PL & 92/92 & $0.4 \pm 0.1$ & $0.61^{+0.07}_{-0.09}$ & $0.5 \pm 0.2$ & $1.0 \pm 0.5$ & \nodata & $<6$ & $8^{+4}_{-2}$ & $2.3^{+0.4}_{-0.3}$ & $2.6^{+1.0}_{-0.7}$\\
& \textbf{\textit{vnei}}+PL & 92/92 & $0.6 \pm 0.2$ & $0.9^{+0.4}_{-0.3}$ & $0.4^{+0.2}_{-0.1}$ & $0.7^{+0.4}_{-0.3}$ & \nodata & $0.17^{+0.18}_{-0.08}$ & $6^{+9}_{-2}$ & $2.4 \pm 0.4$ & $2.7^{+1.1}_{-0.9}$\\
\hline
B & \textit{vpshock}+PL & 59/54 & $0.5 \pm 0.1$ & $0.59^{+0.05}_{-0.08}$ & $0.5 \pm 0.2$ & $0.5 \pm 0.3$ & \nodata & $<12$ & $8^{+4}_{-2}$ & $3.1 \pm 0.4$ & $1.9^{+0.9}_{-0.8}$\\
& \textbf{\textit{vnei}}+PL & 53/54 & $0.4 \pm 0.1$ & $0.60^{+0.04}_{-0.06}$ & $0.6 \pm 0.2$ & $0.6 \pm 0.3$ & \nodata & $2.1^{+1.5}_{-0.7}$ & $6^{+2}_{-1}$ & $3.1 \pm 0.4$ & $1.9^{+0.8}_{-0.6}$\\
\hline
C & \textbf{\textit{vpshock}} & 29/36 & $0.4 \pm 0.1$ & $1.1 \pm 0.3$ & $0.3 \pm 0.1$ & $0.7 \pm 0.3$ & \nodata & $0.5^{0.5}_{-0.2}$ & $2.5^{+1.0}_{-0.5}$ & \nodata & \nodata\\
& \textit{vnei} & 33/36 & $0.31^{+0.10}_{-0.08}$ & $1.36^{+0.08}_{-0.23}$ & $0.3^{+0.2}_{-0.1}$ & $0.7 \pm 0.3$ & \nodata & $0.20^{+0.09}_{-0.10}$ & $1.6^{+0.5}_{-0.3}$ & \nodata & \nodata\\
\hline
D & \textbf{\textit{vpshock}}+PL & 66/66 & $0.41^{+0.19}_{-0.07}$ & $0.37^{+0.04}_{-0.09}$ & \nodata & \nodata & $0.8 \pm 0.1$ & $3^{+7}_{-1}$ & $15^{+46}_{-5}$ & $2.9^{+0.4}_{-0.3}$ & $2.2^{+1.0}_{-0.5}$\\
& \textit{vnei}+PL & 66/66 & $0.4^{+0.2}_{-0.1}$ & $0.26^{+0.02}_{-0.03}$ & \nodata & \nodata & $0.8 \pm 0.1$ & $>0.8$ & $40^{+60}_{-10}$ & $3.0^{+0.3}_{-0.3}$ & $2.5^{+0.6}_{-0.5}$\\
\hline
E & \textit{vpshock} & 38/38 & $0.30 \pm 0.09$ & $0.64^{+0.04}_{-0.03}$ & $0.4^{+0.2}_{-0.1}$ & $0.3 \pm 0.2$ & \nodata & $<16$ & $6 \pm 1$ & \nodata & \nodata\\
& \textbf{\textit{vnei}} & 41/38 & $0.36^{+0.09}_{-0.20}$ & $0.65^{+0.14}_{-0.05}$ & $0.4^{+0.1}_{-0.12}$ & $0.4 \pm 0.2$ & \nodata & $1.2^{+0.6}_{-0.7}$ & $5.4^{+2.0}_{-0.8}$ & \nodata & \nodata \\
\hline
F & \textit{vpshock} & 54/44 & $0.6 \pm 0.1$ & $0.63^{+0.06}_{-0.07}$ & $0.22^{+0.10}_{-0.09}$ & $0.4 \pm 0.2$ & \nodata & $1.8^{+2.7}_{-0.8}$ & $9^{+4}_{-2}$ & \nodata & \nodata\\
& \textbf{\textit{vnei}} & 53/44 & $0.6 \pm 0.2$ & $0.67^{+0.16}_{-0.09}$ & $0.23^{+0.10}_{-0.09}$ & $0.4 \pm 0.2$ & \nodata & $0.7^{+0.3}_{-0.2}$ & $8 \pm 3$ & \nodata & \nodata\\
\enddata
\tablenotetext{a}{Thermal normalization parameter. $N_a=10^{-14}/4 \pi D^2 \int n_e n_H dV$, where $D$ is the distance to the source (50 kpc) and the integral is the volume emission measure.}
\tablenotetext{b}{Power-law normalization parameter, $N_b=1$ photon keV$^{-1}$ cm$^{-2}$ s$^{-1}$ at 1 keV.}
\tablecomments{Errors are 90\% confidence intervals. Elemental abundances are shown relative to solar values. Abundances not shown are fixed to ISM values. The preferred model for each region is bolded.}
\label{tbl:nei_fits}
\end{deluxetable}

\clearpage
\begin{deluxetable}{clccccccc}
\tablecaption{Fits to the hard component in Regions A, B, and D}
\tablehead{
	\colhead{Region} &
	\colhead{Model} &
	\colhead{$\chi^2 / \nu$} &
	\colhead{$\alpha$} &
	\colhead{$\nu_{\text{break}}$} &
	\colhead{$S_{\text{1GHz}}$} &
	\colhead{$E_{\text{max}}$} &
	\colhead{$\Gamma$} &
	\colhead{norm$_{PL}$}
	\\
& & & & ($10^{17}$ Hz) & (mJy) & (TeV\tablenotemark{a}) & & ($10^{-5} N_b$\tablenotemark{b})
}
\startdata
A & srcut & 92/91 & $0.5^{+0.5}_{-0.1}$ & $1.4^{+72}_{-1.0}$ & $<7$ & $30^{+180}_{-14}$ & \nodata & \nodata\\
A & powerlaw & 91/92 & \nodata & \nodata & \nodata & \nodata & $2.5^{+0.5}_{-0.4}$ & $2.7^{+1.3}_{-0.9}$\\
B & srcut & 58/52 & $0.4^{+0.4}_{-0.1}$ & $0.18^{+0.33}_{-0.09}$ & $0.6^{+2.3}_{-0.5}$ & $11^{+7}_{-3}$ & \nodata & \nodata\\
B & powerlaw & 53/54 & \nodata & \nodata & \nodata & \nodata & $3.1 \pm 0.4$ & $1.9^{+0.8}_{-0.6}$\\
D & srcut & 65/65 & $0.5^{+0.3}_{-0.1}$ & $0.3^{+0.8}_{-0.1}$ & $<48$ & $14^{+12}_{-3}$ & \nodata & \nodata\\
D & powerlaw & 66/66 & \nodata & \nodata & \nodata & \nodata & $2.9^{+0.4}_{-0.3}$ & $2.2^{+1.0}_{-0.5}$
\enddata
\tablenotetext{a}{Calculated assuming a magnetic field $B=10$ $ \mu$G}
\tablenotetext{b}{Power-law normalization parameter, $N_b=1$ photon keV$^{-1}$ cm$^{-2}$ s$^{-1}$ at 1 keV.}
\tablecomments{Errors are 90\% confidence intervals.}
\label{tbl:nonthermal_fits}
\end{deluxetable}

\clearpage
\begin{deluxetable}{ccccccc}
\tablecaption{X-ray Spectral Fits for Halo Regions}
\tablehead{
	\colhead{Azimuthal Angle} &
	\colhead{$\chi^2 / \nu$} &
	\colhead{$kT$} &
	\colhead{$n_e t$} &
	\colhead{norm} &
	\colhead{$\Gamma$} &
	\colhead{norm$_{PL}$}\\
	& & (keV) & (10$^{11}$ cm$^{-3}$ s) & 10$^{-5} N_a$\tablenotemark{a} & & 10$^{-5} N_b$\tablenotemark{b}
}
\startdata
0$\degree$ & 130/111 & $0.68^{+0.08}_{-0.05}$ & $>1.6$ & $3.7^{+0.7}_{-0.8}$ & $3.4 \pm 0.2$ & $5.5^{+0.5}_{-1.2}$\\
30$\degree$ & 124/109 & $0.62^{+0.08}_{-0.05}$ & $2.2^{+4.7}_{-0.6}$ & $4.8^{+0.7}_{-1.1}$ & $3.0 \pm 0.2$ & $4.5^{+0.4}_{-0.9}$\\
60$\degree$ & 122/110 & $0.61^{+0.07}_{-0.05}$ & $3^{+10}_{-1}$ & $3.9^{+0.7}_{-0.8}$ & $3.1 \pm 0.2$ & $4.6^{+0.4}_{-0.9}$\\
90$\degree$ & 122/107 & $0.63^{+0.09}_{-0.07}$ & $2.5^{+6.5}_{-0.9}$ & $3.3^{+0.6}_{-1.1}$ & $3.2 \pm 0.2$ & $4.7^{+0.4}_{-0.8}$\\
120$\degree$ & 129/107 & $0.60^{+0.09}_{-0.06}$ & $2.2^{+3.3}_{-0.6}$ & $3.8^{+0.6}_{-1.1}$ & $2.9 \pm 0.2$ & $4.0^{+0.4}_{-0.9}$\\
150$\degree$ & 129/107 & $0.60^{+0.10}_{-0.07}$ & $2.3^{+3.4}_{-0.7}$ & $3.8^{+0.7}_{-1.1}$ & $2.9 \pm 0.2$ & $4.0^{+0.4}_{-0.8}$\\
180$\degree$ & 127/106 & $0.69^{+0.08}_{-0.05}$ & $>2.5$ & $3.7^{+0.4}_{-0.8}$ & $2.5^{+0.2}_{-0.1}$ & $3.1 \pm 0.3$\\
210$\degree$ & 126/107 & $0.78^{+0.07}_{-0.06}$ & $>7$ & $3.7^{+0.4}_{-0.8}$ & $2.5^{+0.2}_{-0.1}$ & $3.1 \pm 0.3$\\
240$\degree$ & 117/109 & $0.74^{+0.06}_{-0.05}$ & $>9$ & $4.7 \pm 0.4$ & $2.5^{+0.2}_{-0.1}$ & $3.1 \pm 0.3$\\
270$\degree$ & 116/109 & $0.74^{+0.06}_{-0.05}$ & $>7$ & $4.7 \pm 0.9$ & $2.6^{+0.2}_{-0.3}$ & $3.5 \pm 0.7$\\
300$\degree$ & 99/109 & $0.75^{+0.06}_{-0.05}$ & $>9$ & $4.8^{+0.4}_{-0.6}$ & $2.7^{+0.2}_{-0.1}$ & $3.7 \pm 0.3$\\
330$\degree$ & 107/111 & $0.72^{+0.06}_{-0.05}$ & $>9$ & $4.8^{+0.5}_{-0.4}$ & $3.0^{+0.2}_{-0.1}$ & $4.6 \pm 0.3$
\enddata
\tablenotetext{a}{Thermal normalization parameter. $N_a=10^{-14}/4 \pi D^2 \int n_e n_H dV$, where $D$ is the distance to the source (50 kpc) and the integral is the volume emission measure.}
\tablenotetext{b}{Power-law normalization parameter, $N_b=1$ photon keV$^{-1}$ cm$^{-2}$ s$^{-1}$ at 1 keV.}
\tablecomments{Errors are 90\% confidence intervals. Elemental abundances are fixed to LMC ISM values.}
\label{tbl:interior_fits}
\end{deluxetable}

\clearpage
\begin{deluxetable}{lccc}
\tablecaption{Shock Speeds and Densities for Shell Regions}
\tablehead{
	\colhead{Region} &
	\colhead{$v$} &
	\colhead{$n_e$ \tablenotemark{a}} &
        \colhead{$n_e$ \tablenotemark{b}} \\
	& (km s$^{-1}$) & (cm$^{-3}$ $v_{10}$) & (cm$^{-3} f^{-1/2}$)
}
\startdata
A & $700 \pm 100$ & $1.0^{+1.4}_{-0.5}$ & $0.5^{+0.3}_{-0.1}$\\
B & $510^{+30}_{-20}$ & $10^{+7}_{-3}$ & $4.0^{+0.6}_{-0.3}$\\
C & $680^{+90}_{-100}$ & $3 \pm 1$ & $2.0^{+0.4}_{-0.3}$\\
D & $400^{+20}_{-50}$ & $30^{+60}_{-10}$ & $4.4^{+4.4}_{-0.8}$\\
E & $530^{+60}_{-40}$ & $3 \pm 1$ & $5.2^{+0.9}_{-0.4}$\\
F & $530^{+50}_{-20}$ & $6^{+3}_{-2}$ & $7 \pm 1$\\
\enddata
\tablenotetext{a}{Estimated density, calculated by dividing the ionization timescale for each region by the approximate time since shock, based on the radial distance of each region. $v_{10}$ is the initial speed of the SNR shock in multiples of 10,000 km s$^{-1}$.}
\tablenotetext{b}{Estimated density, calculated from the thermal normalization parameter using equation \ref{eq:density_from_norm}.}
\label{tbl:shock_speeds}	
\end{deluxetable}

\end{document}